\documentclass[traditabstract]{aa}  
\usepackage{amsmath,amsfonts,amsbsy,mathrsfs,amssymb}
\usepackage{graphicx}
\usepackage{supertabular,booktabs}
\usepackage{longtable}
\usepackage{txfonts}
\usepackage{afterpage}
\usepackage[pdftex=true, colorlinks=true, linkcolor=blue, citecolor=blue, pdfborder={0 0 0}, bookmarks=false, urlcolor=blue, pdftitle={eXtended CASA Line Analysis Software Suite (XCLASS)}, pdfsubject={}, pdfauthor={}, pdfkeywords={}]{hyperref}

\usepackage[numbers,curly,sort]{natbib}
\usepackage{color}
\bibpunct{(}{)}{;}{a}{}{,}          
\begin{document}
    \title{eXtended CASA Line Analysis Software Suite (XCLASS)\thanks{\url{http://www.astro.uni-koeln.de/projects/schilke/myXCLASSInterface}}}


    \author{T.~M\"{o}ller\inst{1}
            \and
            C.~Endres\inst{1,2}
            \and
            P.~Schilke\inst{1}
            }

    \institute{I. Physikalisches Institut, Universit\"{a}t zu K\"{o}ln,
               Z\"{u}lpicher Str. 77, D-50937 K\"{o}ln, Germany\\
               \email{moeller@ph1.uni-koeln.de}
              \and
               Max-Planck-Institut f\"{u}r extraterrestrische Physik, Giessenbachstrasse 1, D-85748 Garching, Germany\\
              }

    \date{Received August 17, 2015 / Accepted September 22, 2016}

    \abstract
{The eXtended CASA Line Analysis Software Suite (XCLASS) is a toolbox for the Common Astronomy Software Applications package (CASA) containing new functions for modeling interferometric and single dish data. Among the tools is the \texttt{myXCLASS} program which calculates synthetic spectra by solving the radiative transfer equation for an isothermal object in one dimension, whereas the finite source size and dust attenuation are considered as well. Molecular data required by the \texttt{myXCLASS} program are taken from an embedded SQLite3 database containing entries from the Cologne Database for Molecular Spectroscopy CDMS) and JPL using the Virtual Atomic and Molecular Data Center (VAMDC) portal. Additionally, the toolbox provides an interface for the model optimizer package Modeling and Analysis Generic Interface for eXternal numerical codes (MAGIX), which helps to find the best description of observational data using \texttt{myXCLASS} (or another external model program), that is, finding the parameter set that most closely reproduces the data.}


    \titlerunning{XCLASS}
    \authorrunning{T.~M\"{o}ller \textit{et al.}}

    \maketitle

    \section{Introduction}\label{sec:Introduction}

The CASA package \cite{McMullin2007} provides a powerful tool for data post-processing for ALMA and JVLA, but contains only rudimentary functions for modeling the data. But modeling astronomical data is essential to derive physical parameters such as column densities and rotational temperatures, as well as information on the location and the kinematics of the emitting gas component for each observed molecular species.

The toolbox described in this paper offers not only the possibility to model data by using the \texttt{myXCLASS} program \cite{Schilke2001, Comito2005, Zernickel2012, Crockett2014a, Crockett2014b, Neill2014}, which computes synthetic spectra by solving the detection equation similar to software packages like Weeds \cite{Maret2011} and CASSIS\footnote{\url{http://cassis.irap.omp.eu/}}, but also makes use of the optimization package \texttt{MAGIX} \citet{Moeller2013}. \texttt{MAGIX} provides a framework of an easy interface between existing codes and an iterating engine that attempts to minimize deviations of the model results from available observational data, constraining the values of the model parameters and providing corresponding error estimates. In addition to the \texttt{myXCLASS} program many external model programs\footnote{Here, the phrase ``external model program'' means the external program that calculates the model function depending on several input parameters.} such as RADEX \cite{vdTak2009}, RADMC-3D\footnote{
\url{http://www.ita.uni-heidelberg.de/~dullemond/software/radmc-3d/}} or LIME \cite{Brinch2010} can be plugged into \texttt{MAGIX} to explore their parameter space and find the set of parameter values that best fits observational data. Most of the optimization algorithms included in the \texttt{MAGIX} package are available as an \texttt{OpenMP}\footnote{\url{http://openmp.org/wp/}} and an \texttt{MPI}\footnote{\url{http://www.open-mpi.org/}} version to speed up the computation. Furthermore, the toolbox includes two new functions which provide a simplified interface for \texttt{MAGIX} in conjunction with \texttt{myXCLASS} to model single spectra and complete data cubes, respectively. Once installed, it is seamlessly integrated into CASA, so the \texttt{tasklist} command within CASA lists all the new functions. In addition, the \texttt{help} command followed by the name of the function provides a short description of the corresponding function with a short overview of all input parameters. In addition, the XCLASS interface can be used without CASA, wherefore the following python packages has to be installed: \texttt{numpy}, \texttt{scipy}, \texttt{pyfits}, \texttt{matplotlib} and \texttt{sqlite3}.

The Splatalogue database\footnote{\url{http://www.splatalogue.net}} is accessible from within the viewer in CASA. We have opted to use our own SQLite3 database based on the VAMDC\footnote{\url{http://www.vamdc.eu}} implementation, mostly because this allows the user to augment the data by providing the partition function between 1.072 and 1000~K at 110 different temperatures in contrast to seven temperatures in the Splatalogue database, which is derived from the traditional JPL\footnote{\url{http://spec.jpl.nasa.gov}} and CDMS\footnote{\url{http://www.cdms.de}} catalogs. It is hoped that the content of Splatalogue and the VAMDC database will at some point be homogenized.

The {\sc XCLASS} interface for CASA is mostly written in \texttt{python}, whereas the \texttt{myXCLASS} program and most of the algorithms included in the \texttt{MAGIX} package are written in Fortran 90. All required Python modules are already included in the CASA package, so no further software package except the GNU compiler\footnote{\url{https://gcc.gnu.org/}} (\texttt{gfortran} and \texttt{gcc}) with \texttt{OpenMP}/\texttt{OpenMPI} extension has to be installed. The {\sc XCLASS} interface for CASA is available for Linux and Mac OS 10.11 (64 bit).

In the next sections we will describe the \texttt{myXCLASS} program and its algorithms, assumptions, and shortcomings in detail. Afterwards, we describe the SQLite3 database used for the \texttt{myXCLASS} program and the related functions provided by the {\sc XCLASS} interface. In the following we briefly present the \texttt{MAGIX} package with a short overview of the included algorithms. Finally, we give a detailed description of the new functions \texttt{myXCLASSFit} and \texttt{myXCLASSMapFit} used for modeling data with \texttt{myXCLASS} and \texttt{MAGIX}.

    \section{myXCLASS}\label{sec:myXCLASS}

The {\sc XCLASS} interface for CASA contains the \texttt{myXCLASS} program, originally developed by P.~Schilke \cite{Schilke2001, Comito2005, Zernickel2012, Crockett2014a, Crockett2014b, Neill2014} based on the GILDAS\footnote{\url{http://iram.fr/IRAMFR/GILDAS/}} package CLASS, which models a spectrum by solving the radiative transfer equation for an isothermal object in one dimension, the detection equation \cite{Strahler2005}. Here, LTE is assumed, that is, the source function is given by the Planck function of an excitation temperature, which does not need to be the physical temperature, but is constant for all transitions. The \texttt{myXCLASS} function is designed to describe line-rich sources which are often dense, so that LTE is a reasonable approximation. Also, a non-LTE description requires collision rates which are available only for a few molecules.


    \begin{figure}[t]
       \centering
       \includegraphics[width=0.4\textwidth]{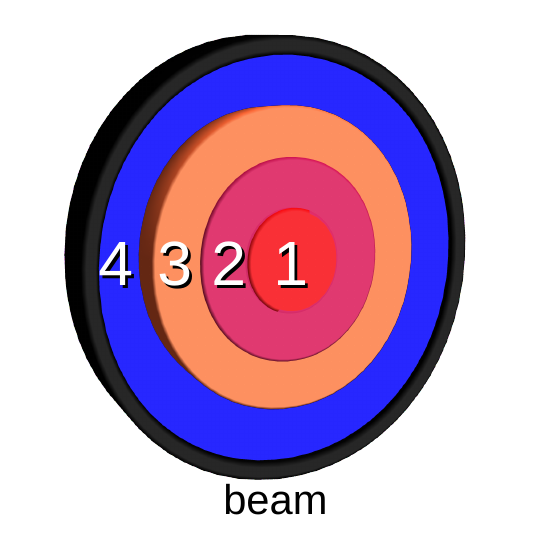}\\
       \caption{Sketch of a distribution of core layers within the Gaussian beam of the telescope (black ring). Here, we assume three different core components $1$, $2$, and $3$, centered at the middle of the beam with different source sizes, excitation temperatures, velocity offsets etc. indicated by different colors. Additionally, we assume that all core components have the same distance to the telescope, i.e.,\ all core layers are located within a plane perpendicular to the line of sight. Furthermore, we assume that this plane is located in front of a background Layer $4$ with homogeneous intensity $I_{\rm bg}^{\rm core}(\nu)$ over the whole beam. Core components do not interact with each other radiatively.}
       \label{fig:myxclassEmissionSketch}
    \end{figure}

The \texttt{myXCLASS} function is able to model a spectrum with an arbitrary number of molecules where the contribution of each molecule is described by an arbitrary number of components. The 1-d assumption imposes a very simplistic geometrical structure. We recognize two classes of components.

One, the core objects (in earlier implementations called emission component), consists of an ensemble of objects centered at the middle of the beam. These could be identified with clumps, hot dense cores etc. which overlaps but do not interact either because they do not overlap in physical or in velocity space. For computational convenience, they are assumed to be centered in the beam, as shown in Fig.~\ref{fig:myxclassEmissionSketch}. It is also assumed that the dust emission emanates (partly) from these components. Their intensities are added, weighted with the beam filling factor, see Eq.~\eqref{myXCLASS:BeamFillingFactor}.

The second class, foreground objects (in earlier implementations called absorption components), are assumed to be in layers in front of the core components. In the current 1-d implementation, they would have a beam averaged intensity of the core sources as background, and would fill the whole beam. Examples for such structures would be source envelopes in front of dense cores, or absorption components along the line-of-sight.


As shown in Fig.~\ref{fig:myxclassEmissionSketch}, we assume that core components do not interact with each other radiatively, that is, one core layer is not influenced by the others. But the core layers may overlap to offer the possibility to model sources consisting of several molecules and compounds. The solution of the radiative transfer equation for core layers is\footnote{A derivation of the expression can be found in the appendix~\ref{deriv:DetectionEquation}.},
\begin{align}\label{myXCLASS:modelEm}
  T_{\rm mb}^{\rm core}(\nu) &= \sum_m \sum_c \Bigg[\eta \left(\theta^{m,c}\right) \left[S^{m,c}(\nu) \left(1 - e^{-\tau_{\rm total}^{m,c}(\nu)} \right) \right.\nonumber\\
                           & \qquad \qquad + \left. I_{\rm bg}^{\rm core} (\nu) \left(e^{-\tau_{\rm total}^{m,c}(\nu)} - 1 \right)\right]\Bigg] \nonumber\\
                           & \quad + \left(I_{\rm bg}^{\rm core} (\nu) -  J_\mathrm{CMB} \right),
\end{align}
where the sums go over the indices $m$ for molecule, and $c$ for (core) component, respectively. In the following we will briefly describe each term in Eq.~\eqref{myXCLASS:modelEm}.\\


The beam filling (dilution) factor $\eta(\theta^{m,c})$ of molecule $m$ and component $c$ in Eq.~\eqref{myXCLASS:modelEm} for a source with a Gaussian brightness profile, see below, and a Gaussian beam is given by\footnote{Derivations of the beam filling factor Eq.~\eqref{myXCLASS:BeamFillingFactor}, are described in the appendix~\ref{deriv:BeamFillingFactor}.}
\begin{equation}\label{myXCLASS:BeamFillingFactor}
    \eta(\theta^{m,c}) = \frac{(\theta^{m,c})^2}{\theta_t(\nu)^2 + (\theta^{m,c})^2},
\end{equation}
where $\theta^{m,c}$ and $\theta_t$ represents the source and telescope beam full width half maximum (FWHM) sizes, respectively. The sources beam FWHM sizes $\theta^{m,c}$ for the different components are defined by the user in the molfit file, described in Sect.~\ref{myxclass:molfit}. Additionally, we assume for single dish observations, that the telescope beam FWHM size is related to the diameter of the telescope by the diffraction limit
\begin{equation}\label{myXCLASS:DiffractionLimit}
  \theta_t(\nu) = \left(1.22 \cdot \frac{\lambda}{D}\right) \cdot \xi = \left(1.22 \cdot \frac{c_{\rm light}}{\nu \, D}\right) \cdot \xi,
\end{equation}
where $D$ describes the diameter of the telescope, $c_{\rm light}$ the speed of light, and $\xi = 3600 \cdot 180 \, \, \pi^{-1}$ a conversion factor to get the telescope beam FWHM size in arcsec. For interferometric observations, the user has to define the interferometric beam FWHM size directly. In contrast to single dish observations we assume a constant interferometric beam FWHM size for the whole frequency range.

The term $\eta_{\rm max}^{\rm core}$ in Eq.~\eqref{myXCLASS:modelEm} indicates the largest beam filling factor of all core components of all molecules, that is, $\displaystyle \eta_{\rm max}^{\rm core} = \max_{m,c} \left\{\eta(\theta^{m,c}) \right\}$.\\


In general, the brightness temperature of radiation temperature $J(T, \nu)$ is defined as
\begin{equation}\label{myXCLASS:JT}
    J(T, \nu) = \frac{h \, \nu}{k_B}\frac{1}{e^{h \, \nu / k \, T} - 1}.
\end{equation}
The expression $J_\mathrm{CMB}$ used in Eq.~\eqref{myXCLASS:modelEm}, describes the radiation temperature Eq.~\eqref{myXCLASS:JT} of the cosmic background $T_{\rm cbg}$~=~2.7~K, that is, $J_\mathrm{CMB} \equiv J(T_{\rm cbg}, \nu)$.\\


In Eq.~\eqref{myXCLASS:modelEm}, the expression $S^{m,c}(\nu)$ represents the source function and is according to Kirchhoff's law of thermal radiation given by
\begin{align}\label{myXCLASS:SourceFunction}
    S^{m,c}(\nu) &= \frac{\epsilon_l^{m,c}(\nu) + \epsilon_d^{m,c}(\nu)}{\kappa_l^{m,c}(\nu) + \kappa_d^{m,c}(\nu)} \nonumber \\
                &= \frac{\kappa_l^{m,c}(\nu) \, J(T_\mathrm{ex}^{m,c}, \nu) + \kappa_d^{m,c}(\nu) \, J(T_d^{m,c}, \nu)}{\kappa_l^{m,c}(\nu) + \kappa_d^{m,c}(\nu)}\nonumber \\
                &= \left(1 - \delta_{\gamma, 0}\right) \cdot \left[\frac{\tau_l^{m,c}(\nu) \, J(T_\mathrm{ex}^{m,c}, \nu) + \tau_d^{m,c}(\nu) \, J(T_d^{m,c}, \nu)}{\tau_l^{m,c}(\nu) + \tau_d^{m,c}(\nu)}\right]\nonumber \\
                & \quad + \delta_{\gamma, 0} \, J(T_\mathrm{ex}^{m,c}, \nu),
\end{align}
where $\epsilon_{l,d}^{m,c}(\nu)$ and $\kappa_{l,d}^{m,c}(\nu)$ are the core and foreground coefficients for line and dust, respectively. Additionally, the optical depth is given by $\tau^{m,c}(\nu) = \int \kappa^{m,c}(\nu) \, ds = \kappa^{m,c}(\nu) \, s$. This assumes that molecules and dust are well mixed, that is, it would not be correct if the molecule exists only in part of the cloud, but the dust everywhere. In older versions, the background temperature could only be defined as the measured continuum offset, which corresponds to the beam-averaged continuum brightness temperature. At the same time, the dust, as agent of line attenuation, was described by column density and opacity. This is practical, because the observable $T_{\rm bg}$ is used, but does not constitute a self-consistent and fully physical description. Therefore, we now use optionally either a physical ($\gamma \equiv 1$) or phenomenological ($\gamma \equiv 0$) description of the background indicated by the Kronecker delta $\delta_{\gamma, 0}$, that is, $S^{m,c}(\nu) \equiv J(T_\mathrm{ex}^{m,c}, \nu)$ for $\gamma \equiv 0$. (Here, the phrase ``background'' means the ``layer'' with intensity $I_{\rm bg}^{\rm core}(\nu)$ which is located behind the core components, that is, the background of the core layers, see Fig.~\ref{fig:myxclassEmissionSketch}.) We note that, if $\gamma \equiv 0$, the definition of the dust temperature~$T_d^{m,c} (\nu)$, Eq.~\eqref{myXCLASS:EffDustTempTbgLocal}, is superfluous.\\


The total optical depth $\tau_{\rm total}^{m,c}(\nu)$ of each molecule $m$ and component $c$ is defined as the sum of the optical depths $\tau_l^{m,c}(\nu)$ of all lines of each molecule $m$ and component $c$ plus the dust optical depth $\tau_d^{m,c}(\nu)$, that is,
\begin{equation}\label{myXCLASS:taud}
    \tau_{\rm total}^{m,c}(\nu) = \tau_l^{m,c}(\nu) + \tau_d^{m,c}(\nu),
\end{equation}


where the dust optical depth $\tau_d^{m,c}(\nu)$ takes the dust attenuation into account and is given by
\begin{align}\label{myXCLASS:dustOpacity}
  \tau_d^{m,c}(\nu) &= \tau_{d, {\rm ref}}^{m,c} \cdot \left(\frac{\nu}{\nu_{\rm ref}} \right)^{\beta^{m,c}} \nonumber\\
                    &= \left(N_H^{m,c} \cdot \kappa^{m,c}_{\nu_{\rm ref}} \cdot m_{H_2} \cdot \frac{1}{\zeta_{\rm gas-dust}}\right) \cdot \left(\frac{\nu}{\nu_{\rm ref}} \right)^{\beta^{m,c}}.
\end{align}
Here, $N_H^{m,c}$ describes the hydrogen column density, $\kappa^{m,c}_{\nu_{\rm ref}}$ the dust mass opacity for a certain type of dust \cite{OssenkopfHenning1994} at the reference frequency $\nu_{\rm ref}$, and $\beta^{m,c}$ the spectral index of $\kappa^{m,c}_{\nu_{\rm ref}}$. These parameters are defined by the user, see Sect.~\ref{myxclass:molfit}. In addition, $\nu_{\rm ref}$ = 230~GHz indicates the reference frequency of the reference dust opacity $\tau_{d, {\rm ref}}^{m,c}$, $m_{H_2}$ describes the mass of a hydrogen molecule, and $\zeta_{\rm gas-dust}^{-1}$ describes the dust to gas ratio and is set here to (1/100) \cite{Hillebrand1983}. The equation is valid for dust and gas well mixed.\\


The optical depth $\tau_l^{m,c}(\nu)$ of all lines for each molecule $m$ and component $c$ is described as\footnote{A derivation of the expression is given in the appendix~\ref{deriv:OpticalDepth}.}
\begin{align}\label{myXCLASS:tau}
    \tau_l^{m,c}(\nu) &= \sum_t \Bigg[\frac{c_{\rm light}^2}{8 \pi \nu^2} \, A_{ul}^t \, N_{\rm tot}^{m,c} \, \frac{g_u^t \, e^{-E_l^t/k_B \, T_{\rm ex}^{m,c}}}{Q \left(m, T_{\rm ex}^{m,c} \right)} \left(1 - e^{-h \, \nu^t /k_B \, T_{\rm ex}^{m,c}} \right) \nonumber\\
                      & \qquad \quad \times \phi^{m,c,t}(\nu)\Bigg],
\end{align}
where the sum with index $t$ runs over all spectral line transitions of molecule $m$ within the given frequency range. The Einstein $A_{ul}$ coefficient\footnote{The indices $u$ and $l$ represent upper and lower state of transition $t$, respectively.}, the energy of the lower state $E_l$, the upper state degeneracy $g_u$, and the partition function $Q \left(m, T_{\rm ex}^{m,c} \right)$ of molecule $m$ are taken from the embedded SQLite3 database, described in Sect.~\ref{sec:db}. (Because the database usually does not describe the partition functions at the given excitation temperature $T_{\rm ex}^{m,c}$, the value of $Q \left(m,T_{\rm ex}^{m,c} \right)$ is computed from a linear interpolation. With the new catalog, extrapolation should not be necessary for most conditions encountered in molecular cores.) In addition, the values of the excitation temperatures $T_{\rm ex}^{m,c}$ and the column densities $N_{\rm tot}^{m,c}$ for the different components and molecules are taken from the user defined molfit file, see Sect.~\ref{myxclass:molfit}.\\

In order to take broadening of lines caused by the thermal motion of the gas particles and micro-turbulence into account we assume in Eq.~\eqref{myXCLASS:tau} a normalized Gaussian line profile, that is, $\int_0^{\infty} \phi(\nu) \, d\nu$ = 1, for a spectral line $t$:
\begin{equation}
    \phi^{m,c,t}(\nu) = \frac{1}{\sqrt{2 \pi} \, \sigma^{m,c,t}} \cdot e^{-\frac{\left(\nu - \left( \nu^t + \delta \nu_{\rm LSR}^{m,c,t} \right) \right)^2} {2 (\sigma^{m,c,t})^2}}.
\end{equation}
The source frequency $\delta \nu_{\rm LSR}^{m,c,t}$ for each component $c$ of a molecule $m$ is related to the user defined velocity offset $\left(\delta {\rm v}_{\rm offset}^{m,c}\right)$ taken from the aforementioned molfit file, by the following expression
\begin{equation}\label{myXCLASS:SourceVelocity}
    \delta \nu_{\rm LSR}^{m,c,t} = -\frac{\delta {\rm v}_{\rm offset}^{m,c}}{c_{\rm light}} \cdot \nu^t,
\end{equation}
where $\nu^t$ indicates the frequency of transition $t$ taken from the SQLite3 database mentioned above. Additionally, the standard deviation~$\sigma^{m,c}$ of the profile is defined by the velocity width $\left(\Delta {\rm v}_{\rm width}^{m,c}\right)$ described in the molfit file for each component $c$ of a molecule $m$:
\begin{equation}\label{myXCLASS:sigma}
    \sigma^{m,c,t} = \frac{\frac{\Delta {\rm v}_{\rm width}^{m,c}}{c_{\rm light}} \cdot \left(\nu^t + \delta \nu_{\rm LSR}^{m,c,t} \right)}{2 \, \sqrt{2 \, \ln 2}}.
\end{equation}


The beam-averaged continuum background temperature $I_{\rm bg}^{\rm core} (\nu)$ is parametrized as
\begin{equation}\label{myXCLASS:UserTbg}
    I_{\rm bg}^{\rm core} (\nu) = T_{\rm bg} \cdot \left(\frac{\nu}{\nu_{\rm min}} \right)^{T_{\rm slope}} + J_\mathrm{CMB}
\end{equation}
to allow the user to define the continuum contribution for each frequency range, individually. Here, $\nu_{\rm min}$ indicates the lowest frequency of a given frequency range. $T_{\rm bg}$ and $T_{\rm slope}$, defined by the user, describe the background continuum temperature and the temperature slope, respectively. Here, the treatment of the dust is not entirely self-consistent. To amend that, we would need to define the sub-beam scale structure of the source, which we consider to be outside the scope of the current effort, although it is envisioned to provide this as an option in the future.\\

In Eq.~\eqref{myXCLASS:modelEm}, the continuum contribution is described through the source function $S^{m,c}(\nu)$, Eq.~\eqref{myXCLASS:SourceFunction}, by an effective dust temperature $T_d^{m,c}$ (through $J(T_d^{m,c}, \nu)$) for each component which is given by
\begin{align}\label{myXCLASS:EffDustTempTbgLocal}
    T_d^{m,c} (\nu) &= T_{\rm ex}^{m,c} (\nu) + \Delta T_d^{m,c}(\nu) \nonumber\\
                    &= T_{\rm ex}^{m,c} (\nu) + T_{\rm d, off}^{m,c} \cdot \left(\frac{\nu}{\nu_{\rm min}} \right)^{T_{\rm d, slope}^{m,c}},
\end{align}
where $T_{\rm d, off}^{m,c}$ and $T_{\rm d, slope}^{m,c}$ can be defined by the user for each component in the molfit. If $T_{\rm d, off}^{m,c}$ and $T_{\rm d, slope}^{m,c}$ are not defined for a certain component, we assume $T_d^{m,c} (\nu) \equiv T_{\rm ex}^{m,c} (\nu)$ for all components. For a physical ($\gamma \equiv 1$) description of the background intensity, see Eq.~\eqref{myXCLASS:SourceFunction}, the user can define the dust opacity, Eq.~\eqref{myXCLASS:dustOpacity}, and dust temperature, Eq.~\eqref{myXCLASS:EffDustTempTbgLocal}, for each component.\\

Finally, the last term $J_\mathrm{CMB}$ in Eq.~\eqref{myXCLASS:modelEm} describes the OFF position for single dish observations, where we have an intensity caused by the cosmic background $J_\mathrm{CMB}$. For interferometric observations, the contribution of the cosmic background is filtered out and has to be subtracted as well.\\


    \begin{figure}[t]
       \centering
       \includegraphics[width=0.48\textwidth]{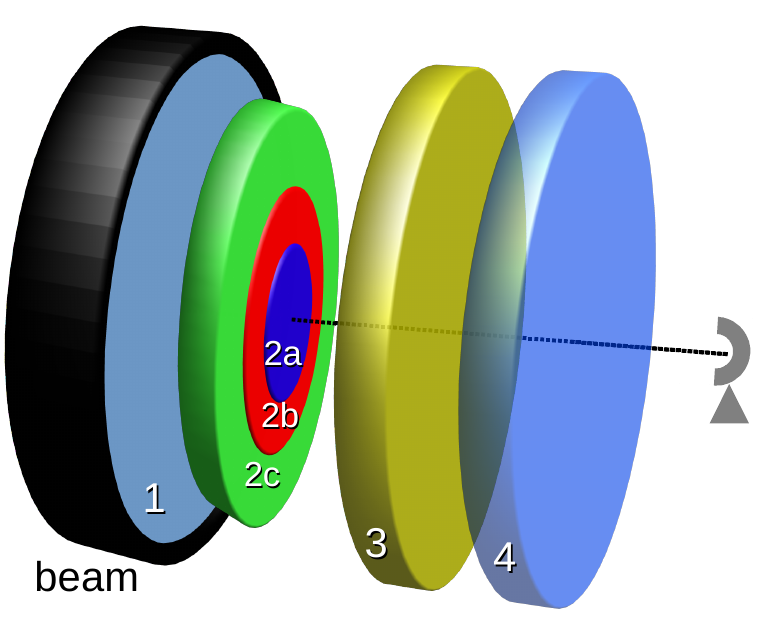}\\
       \caption{Sketch of a distribution of core and foreground layers within the Gaussian shaped beam of the telescope (black ring). Here, we assume three different core components $2a$, $2b$, and $2c$ located in a plane perpendicular to the line of sight which lies in front of the background layer $1$ with intensity $I_{\rm bg}^{\rm core}(\nu)$, see Eq.~\eqref{myXCLASS:UserTbg}. The foreground layers $3$ and $4$ are located between the core layers and the telescope along the line of sight (black dashed line). Here, each component is described by different excitation temperatures, velocity offsets etc. indicated by different colors. The thickness of each layer is described indirectly by the total column density $N_{\rm tot}^{m,c}$, see appendix~\ref{deriv:OpticalDepth}. For each foreground layer we assume a beam filling factor of one, i.e.,\ each foreground layer covers the whole beam.}
       \label{fig:myxclassAbsorptionSketch}
    \end{figure}

In contrast to core layers, foreground components may interact with each other, as shown in Fig.~\ref{fig:myxclassAbsorptionSketch}, where absorption takes places only, if the excitation temperature for the absorbing layer is lower than the temperature of the background.

Hence, the solution of the radiative transfer equation for foreground layers can not be given in a form similar to Eq.~\eqref{myXCLASS:modelEm}. Foreground components have to be considered in an iterative manner. The solution of the radiative transfer equation for foreground layers can be expressed as
\begin{align}\label{myXCLASS:modelAbs}
    T_{\rm mb}^{\rm fore}(\nu)_{m,c=1} &= \eta \left(\theta^{m,c=1} \right) \left(S^{m,c=1}(\nu) - T_{\rm mb}^{\rm core}(\nu)\right) \left(1 - e^{-\tau_{\rm total}^{m,c=1}(\nu)}\right) \nonumber\\
                                      & \quad + T_{\rm mb}^{\rm core}(\nu) \nonumber\\
    T_{\rm mb}^{\rm fore}(\nu)_{m,c=i} &= \eta \left(\theta^{m,c=i} \right) \left(S^{m,c=i}(\nu) - T_{\rm mb}^{\rm fore}(\nu)_{m,c=(i-1)}\right) \nonumber\\
                                      & \quad \times \left(1 - e^{-\tau_{\rm total}^{m,c=i}(\nu)}\right) + T_{\rm mb}^{\rm fore}(\nu)_{m,c=(i-1)},
\end{align}
where $m$ indicates the index of the current molecule and $i$ represents an index running over all foreground components $c$ of all molecules. Additionally, we assume that each foreground component covers the whole beam, that is, $\eta \left(\theta^{m,c=1} \right) \equiv 1$ for all foreground layer. Thus, Eq.~\eqref{myXCLASS:modelAbs} simplifies to
\begin{align}\label{myXCLASS:modelAbsFinal}
    T_{\rm mb}^{\rm fore}(\nu)_{m,c=1} &= \Big[ S^{m,c=1}(\nu) \left(1 - e^{-\tau_{\rm total}^{m,c=1}(\nu)}\right) + T_{\rm mb}^{\rm core}(\nu)  e^{-\tau_{\rm total}^{m,c=1}(\nu)}\Big] \nonumber\\
    T_{\rm mb}^{\rm fore}(\nu)_{m,c=i} &= \Big[S^{m,c=i}(\nu) \left(1 - e^{-\tau_{\rm total}^{m,c=i}(\nu)}\right) \nonumber\\
                                      & \qquad + T_{\rm mb}^{\rm fore}(\nu)_{m,c=(i-1)} e^{-\tau_{\rm total}^{m,c=i}(\nu)}\Big],
\end{align}
where $T_{\rm mb}^{\rm core}(\nu)$ describes the core spectrum, see Eq.~\eqref{myXCLASS:modelEm}, including the beam-averaged continuum background temperature $I_{\rm bg}^{\rm core} (\nu)$. For foreground lines the contribution by other components is considered by first calculating the contribution of core objects and then use this as new continuum for foreground lines reflecting the fact that cold foreground layers are often found in front of hotter emission sources. We assume, that the cosmic background describes together with the core components one end of a stack of layers. Additionally, the foreground components are located between this plane and the telescope, see Fig.~\ref{fig:myxclassAbsorptionSketch}. The total column density $N_{\rm tot}^{m,c}$ depends on the abundance of a certain molecule and on the thickness of a layer containing the molecule. The order of components along the line of sight is defined by the occurrence of a certain foreground component in the molfit file.\\

By fitting all species and their components at once, line blending and optical depth effects are taken into account. The modeling can be done simultaneously with isotopologues (and higher vibrational states) of a molecule assuming an isotopic ratio stored in the so-called iso ratio file, see Sect.~\ref{myxclass:iso}. Here, all parameters are expected to be the same except the column density which is scaled by one over the isotopic ratio for each isotopologue.\\

    \begin{figure}[t]
       \centering
       \includegraphics[width=0.495\textwidth]{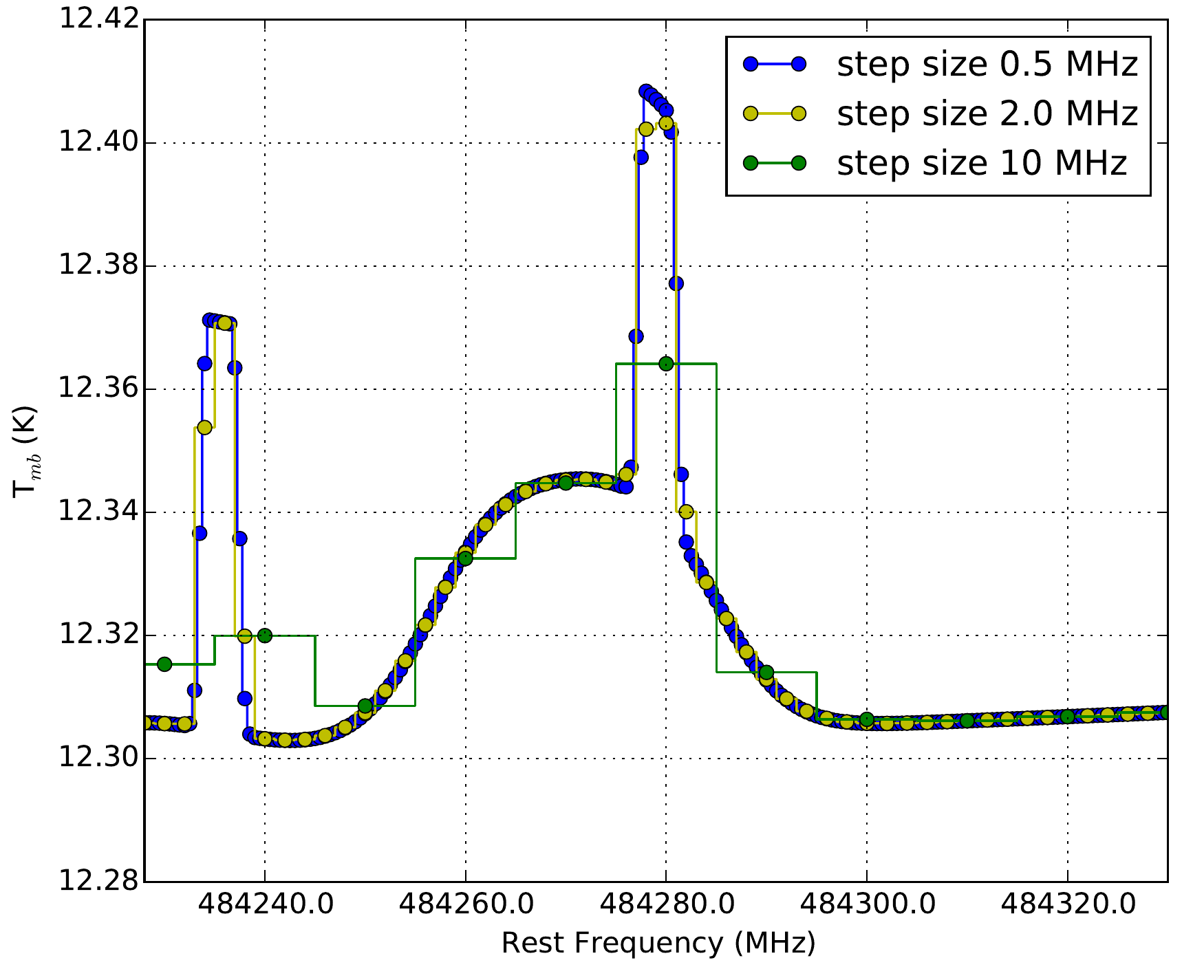}\\
       \caption{Here, the \texttt{myXCLASS} function was used to calculate a spectrum for three different step sizes.}
       \label{fig:myxclassInt}
    \end{figure}

In order to correctly take instrumental resolution effects into account in comparing the modeled spectrum with observations \texttt{myXCLASS} integrates the calculated spectrum over each channel. Thereby, \texttt{myXCLASS} assumes that the given frequencies $\nu$ describe the center of each channel, respectively. This is particularly important if the instrumental channel width is in the order of, or even larger, than intrinsic line widths. The resulting value is than given as
\begin{equation}\label{myXCLASS:Int}
  T_{\rm mb}(\nu) = \frac{1}{\Delta \nu_c}\int_{\nu-\frac{\Delta \nu_c}{2}}^{\nu+\frac{\Delta \nu_c}{2}} T_{\rm mb}(\tilde{\nu})d\tilde{\nu},
\end{equation}
where $\Delta \nu_c$ represents the width of a channel. Due to the complexity of Eqs.~\eqref{myXCLASS:modelEm} and \eqref{myXCLASS:modelAbsFinal} the integration in Eq.~\eqref{myXCLASS:Int} can not be done analytically. Therefore, \texttt{myXCLASS} performs a piecewise integration of each component and channel using the trapezoidal rule and sums up the resulting values to get the final value used in Eq.~\eqref{myXCLASS:Int}. Internal resampling guarantees that even if the line width is smaller than the channel width the integration is done correctly, see Fig.~\ref{fig:myxclassInt}.

\subsection{The molfit file}\label{myxclass:molfit}

Within the molfit file the user defines both which molecules are taken into account and the number of components for each molecule. Additionally, the user has to define for each component the source size $\theta^{m,c}$ in arcsec (\texttt{size}), the excitation temperature $T_{\rm ex}$ in K (\texttt{T\_ex}), the column density $N_{\rm tot}$ in cm$^{-2}$ (\texttt{N\_tot}), the velocity width (FWHM) $\Delta \nu$ in km s$^{-1}$ (\texttt{V\_width}), the velocity offset in km s$^{-1}$ (\texttt{V\_off}), which is via Eq.~\eqref{myXCLASS:SourceVelocity} connected to the source velocity $\nu_{\rm LSR}$, and the flag (\texttt{CFFlag}) indicating if a component is considered for core \texttt{c} or foreground \texttt{f}. The definition of the source size $\theta^{m,c}$ for an foreground component will be ignored, because we assume that all foreground layers covers the whole beam. So, the definition of this parameter is not necessary.\\

\noindent Example of a molfit file:
\begin{verbatim}
% Number of molecules = 2
% size:  T_ex:    N_tot: V_width: V_off:  CFFlag:
CS;v=0;   3
  48.47  80.00  3.91E+17     2.86 -20.56  c
  21.80  51.03  6.96E+17     8.07  30.68  c
  81.70  68.11  1.46E+17     5.16 -10.12  c
HCS+;v=0;   2
        150.00  1.10E+18     5.00  -0.15  f
        200.00  2.20E+17     3.10  -2.15  f
\end{verbatim}

The definition of parameters for a molecule starts with a line describing the name of the molecule, which must be identical to the name of the molecule included in the database, see Sect.~\ref{sec:db}, followed by the number of components $N$ for this molecule. The following $N$ lines describe the parameters for each components, separately. Generally, all parameters have to be separated by blanks, comments are marked with the \% character.\\


In order to define the dust temperature for each component, the molfit file has to contain two additional columns between columns \texttt{V\_off} and \texttt{CFFlag}, describing the parameters $T_{\rm d, off}^{m,c}$ and $T_{\rm d, slope}^{m,c}$, Eq.~\eqref{myXCLASS:EffDustTempTbgLocal}, respectively.\\


Additionally, the \texttt{myXCLASS} program allows to define a hydrogen column density $N_H$ (in cm$^{-2}$), dust mass opacity $\kappa_{\nu_{\rm ref}}$ (in cm$^2$ g$^{-1}$), and the spectral index $\beta$ for each component or globally, that is, $N_H^{m,c} \rightarrow N_H$, $\kappa^{m,c}_{\nu_{\rm ref}} \rightarrow \kappa_{\nu_{\rm ref}}$, and $\beta^{m,c} \rightarrow \beta$. In order to define these parameters individually for each component, the molfit file has to contain additional columns on the left side of column \texttt{CFFlag}.

For globally defined dust parameters, the \texttt{myXCLASS} program assumes that all core components do not contain dust except the core component with the largest beam filling factor, see Eq.~\eqref{myXCLASS:BeamFillingFactor}. This avoids an overestimation of the dust contribution caused by the overlap of the core components.

\subsection{The iso ratio file}\label{myxclass:iso}

As mentioned above, the \texttt{myXCLASS} program offers the possibility to define isotopologues of a molecule and their abundance ratios to reduce the number of input parameters. For that purpose the user has to create an iso ratio file which consists of three columns, where the first two columns indicates the isotopologue and the corresponding molecule, respectively. The third column defines the assumed isotopic ratio. The columns are separated by blanks and comments are marked with the \% character.


\noindent Example of an iso ratio file:
\begin{verbatim}
  % isotopologue:   molecule:   ratio:
  S-33-O2;v=0;      SO2;v=0;      30.0
  S-34-O2;v=0;      SO2;v=0;       5.0
  SO2;v2=1;         SO2;v=0;       1.0
  SOO-17;v=0;       SO2;v=0;     750.0
  SOO-18;v=0;       SO2;v=0;     500.0
\end{verbatim}

Here, we assume that the abundance of molecule \texttt{"SO2;v=0;"} is 30 times higher than the abundance of its isotopologue \texttt{"S-33-O2;v=0;"}.

\subsection{The myXCLASS function}\label{myxclass:func}

\begin{figure}[t]
       \centering
       \includegraphics[width=0.53\textwidth]{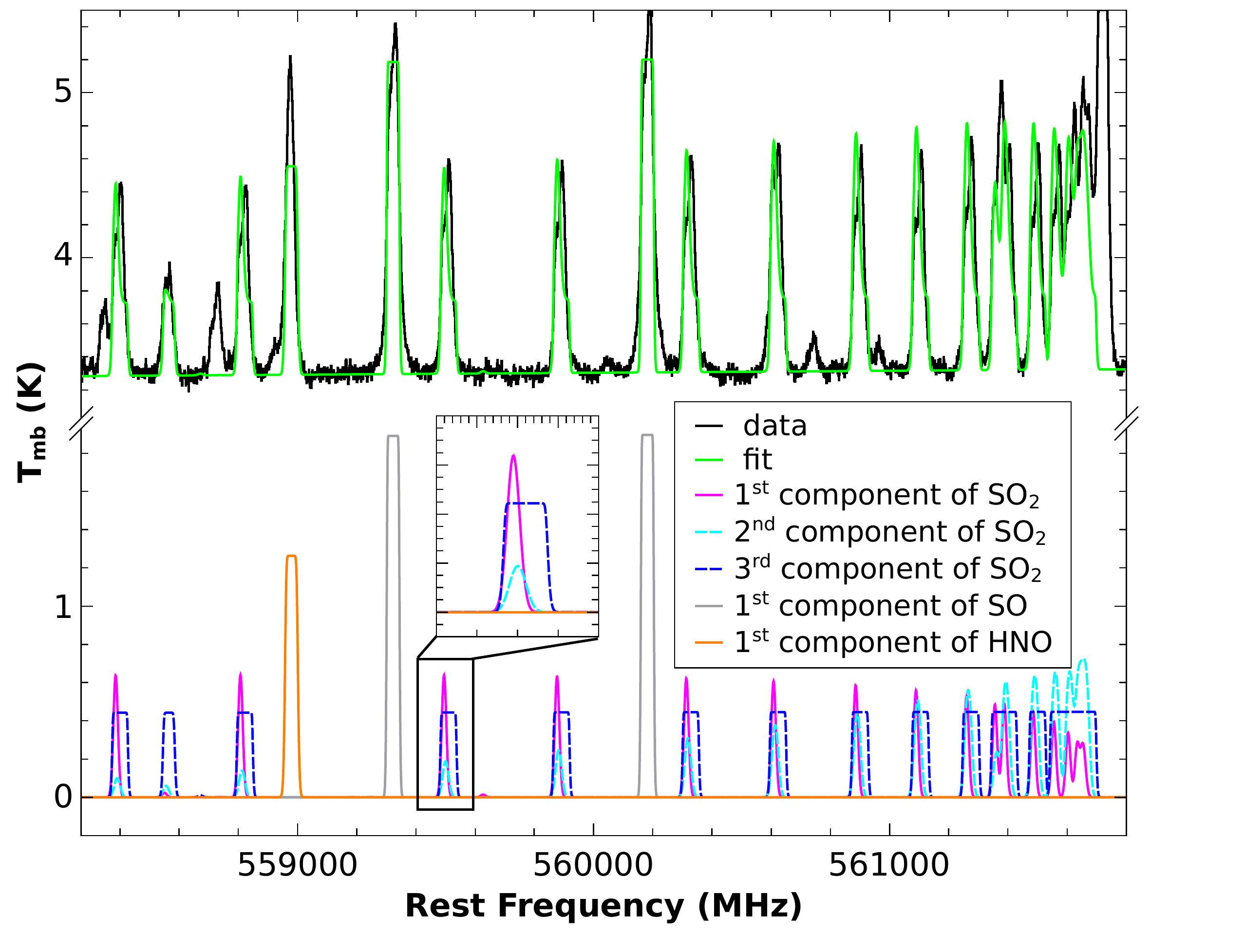}\\
       \caption{Here, the \texttt{myXCLASS} function was used to model HIFI data of Sgr B2m (black) using SO$_2$ (with three different components), SO (with one component), and HNO (with one component). The intensities of each component are shown in the bottom half.}
       \label{fig:myxclass}
    \end{figure}

The \texttt{myXCLASS} function calculates a synthetic spectrum for a user defined frequency range, see Fig.~\ref{fig:myxclass}. The function returns the calculated \texttt{myXCLASS} spectrum, a list of all transition frequencies within the defined range, and the intensities and optical depths for each molecule and component as python arrays, which can be used for further analysis. An example call of the \texttt{myXCLASS} function is given in appendix~\ref{app:myxclass}.
\vspace{.5cm}

    \section{Database}\label{sec:db}

The {\sc XCLASS} interface contains a SQLite3 database including spectroscopic data from CDMS \cite{Mueller2001, Mueller2005} and JPL \citet{Pickett1}. Data can be retrieved and updated via VAMDC, which is an interoperable e-Infrastructure for the exchange of molecular and atomic data between different databases. In principle, all databases which support the VAMDC standard and which contain the required spectroscopic information can be accessed via \texttt{myXCLASS}. A local database was chosen to be able to run the program without network access. CDMS also provides the latest version of the SQLite3 database, which is updated when new entries to the CDMS or JPL are made. In addition, the {\sc XCLASS} interface allows in principle to add private entries (e.g.,\ unpublished new spectroscopic data) to the database. New functions which facilitate adding private entries will be included in one of the next releases of the {\sc XCLASS} package.\\

\noindent The spectroscopic data is stored in two tables: The table \texttt{partitionfunctions} contains the partition functions $Q (m, T_{\rm ex})$ for more than 1000 molecules for 110 different temperatures between 1.072 and 1000 K, which widely extends the standard range of the CDMS/JPL entries. This feature was deemed desirable because particularly for foreground lines with an excitation temperature $T_{\rm ex}$ of 2.7~K, extrapolation from the canonical values (lowest calculated temperature of the partition function 9.75 K) could be wrong in either direction by large factors. Additonally, the table \texttt{transitions} contains more than 5 415 000 transitions and includes the frequency $\nu^t$ (in MHz) with its uncertainty, the Einstein $A_{ul}$ coefficient (in $s^{-1}$), the upper state degeneracy $g_{u}$, the energy of the lower state $E_l$ (in K), and the quantum numbers for each transition.\\

\noindent Entries in the tables are related by the species name, which are different from the names used in the Splatalogue catalog, for example, we use \texttt{"HC-13-N;v=0;"} instead of \texttt{"HC(13)N v=0"} (Splatalogue). In one of the next releases the Splatalogue names will be usable as well.\\

\begin{figure}[t]
      \centering
      \includegraphics[width=0.47\textwidth]{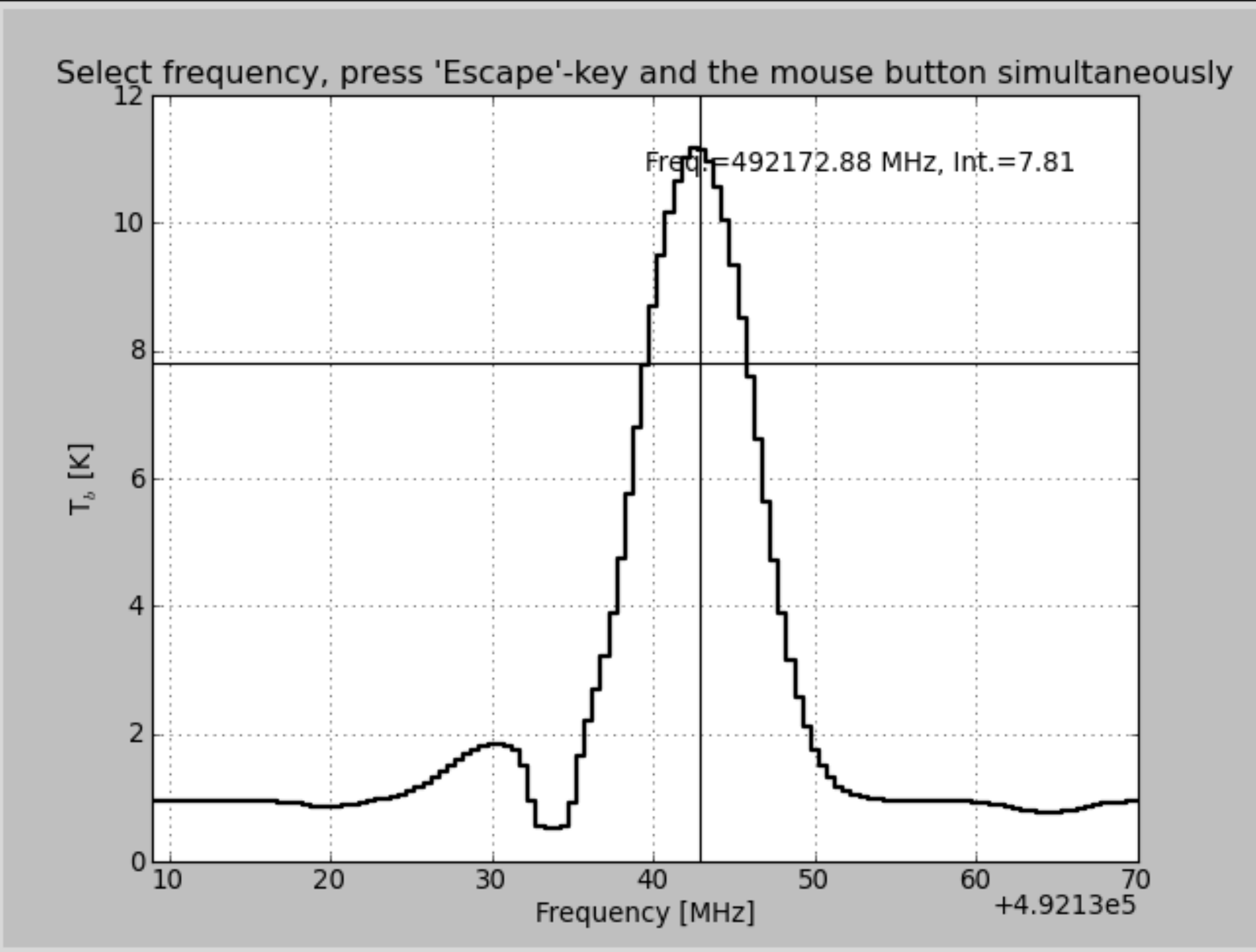}\\
      \caption{Graphical user interface (GUI) of the \texttt{GetTransitions} function. Here, the user has to define a frequency range by selecting a single frequency in the shown spectrum. This frequency represents the central frequency of the range. The width of the range is given by an user defined parameter. XCLASS prints out informations (name of molecule, transition frequency, uncertainty of transition frequency, Einstein A coefficient, quantum number for lower and upper state, and energy of lower state) of all transitions in the selected molecule list located in the given frequency range, see screen output of \texttt{ListDatabase} described in appendix~\ref{app:listdatabase}.}
      \label{fig:GetTransition}
    \end{figure}

Furthermore, the {\sc XCLASS} interface provides functions to update and manage the \texttt{myXCLASS} database: The function \texttt{UpdateDatabase} updates the existing database file by downloading the latest changes from the databases via VAMDC or by downloading a complete new database file from the CDMS server. The first option adds new entries to the tables and overwrites existing entries with new data. Private entries remain unchanged. The second option is much faster but removes all private entries by overwriting the database file with the most recent file from the CDMS server. In addition the function \texttt{DatabaseQuery} offers the possibility to send an user defined SQL query string to the database, for example, to add private entries or query about species names, content etc. Also, the interface contains functions \texttt{ListDatabase}, see appendix~\ref{app:listdatabase}, and \texttt{GetTransitions}, see Fig.~\ref{fig:GetTransition}, to display information about transitions within an user defined range. The \texttt{GetTransitions} function allows the user to select a region on the displayed spectrum, and prints out all the lines from the catalog around the region selected. Additionally, the function offers the possibility to set filters in frequency and energy range.

    \section{MAGIX}\label{sec:MAGIX}

So far, we have discussed how to use XCLASS to produce synthetic spectra. In fact, what the user usually desires is a description of data that provides a fit to a set of observational data. In many packages, this is done by manually changing the parameters until the fit looks good. In XCLASS, we use an optimizer that provides the parameter set for a given model which best matches the data. Due to the large number of input parameters required by the \texttt{myXCLASS} program it is essential to use a powerful optimization package to achieve a good description of observational data by the \texttt{myXCLASS} program. Therefore, the {\sc XCLASS} interface contains the \texttt{MAGIX} package\citet{Moeller2013} which is a model optimizer providing an interface between existing codes and an iterating engine. The package attempts to minimize deviations of the model results from observational data, constraining the values of the model parameters and providing corresponding error estimates.

\begin{figure}[t]
       \centering
        \includegraphics[width=0.38\textwidth]{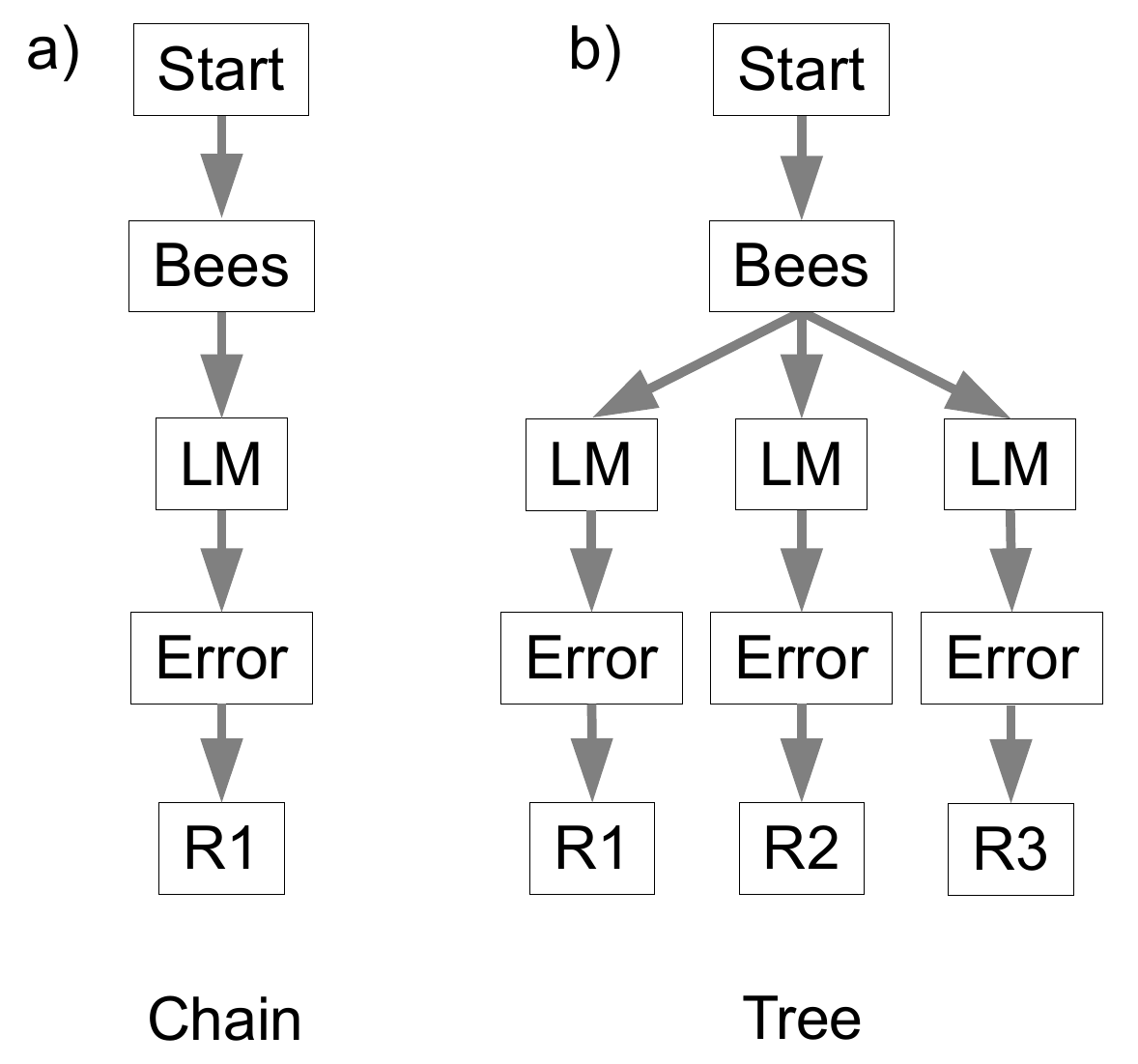}\\
       \caption{Example of an a) algorithm chain and b) algorithm tree. In both cases, we start with the bees swarm algorithm and use the best (a) and the three best results (b) as starting values for the Levenberg-Marquardt local optimization algorithm (LM), respectively. Afterwards, we apply the error estimation algorithm (Error) to the result(s) of the Levenberg-Marquardt algorithm. In case of an algorithm chain we get one final result (R1), whereas we get three different results (R1, R2, R3) for an algorithm tree.}
       \label{plot:AlgTree}
     \end{figure}

\texttt{MAGIX} offers the possibility to model physical and chemical data using an arbitrary external model program, not only \texttt{myXCLASS}, explore their parameter space and find the set of parameter values that best fits observational data. The goodness of a fit is described by the $\chi^2$ distribution which is a function of relative quadratic differences between observational and model values. \texttt{MAGIX} is able to explore the landscape of the $\chi^2$ function without the knowledge of starting values. Additionally, \texttt{MAGIX} can calculate probabilities for the occurrence of minima in the $\chi^2$ distribution and give information about confidence intervals for the parameters. The package provides optimization through one of the following algorithms or via a combination of several of them (an algorithm chain or tree, see Fig.~\ref{plot:AlgTree}): the Levenberg-Marquardt (conjugate gradient) method, which is fast, but can get stuck in local minima, simulated annealing, which is slower, but more robust against local minima. Other global optimization algorithms, such as bees, genetic, particle swarm optimization, nested sampling, or interval nested sampling algorithms are included as well for exploring the solution landscape, checking for the existence of multiple solutions, and giving confidence ranges. Additionally, \texttt{MAGIX} provides an interface to make several algorithms included in the \texttt{scipy}\footnote{\url{http://www.scipy.org/}} package available.

Using an algorithm chain or tree offers the possibility to send the results of the optimization process performed by one algorithm to another optimization loop through some different algorithm. The simulated annealing as well as the Levenberg-Marquardt algorithm require starting values of the parameters that are not too far from the final solution, otherwise the algorithm might stray so far away from the solution that it never converges, that is, the user has to find an acceptable fit by hand before applying these algorithms produces useful results. Often, the location of the minimum can be guessed with sufficient accuracy to give good starting values, but sometimes one is completely in the dark. Using an algorithm chain or tree, the user can first apply one of the swarm algorithms, for example, the bees or interval nested sampling algorithm, to determine the starting values for the subsequent local optimization algorithm using simulated annealing or the Levenberg-Marquardt algorithm, see Fig.~\ref{plot:AlgTree}a). But \texttt{MAGIX} does not only allow one to use the best but also the second best etc. result of a swarm algorithm as starting values for other algorithms (algorithm tree), see Fig.~\ref{plot:AlgTree}b). Therefore, \texttt{MAGIX} is able to find multiple minima of models.

Since this can, depending on the data set and the number of free parameters, be quite computing intensive, \texttt{OpenMP} and \texttt{OpenMPI} parallelization for simultaneous evaluations of the external program is available for most of the algorithms. In addition to the normal \texttt{MAGIX} package which can be used with an arbitrary external model program, the {\sc XCLASS} interface contains a further \texttt{MAGIX} version optimized for the usage with the \texttt{myXCLASS} program. This optimized \texttt{MAGIX} version is used by the \texttt{myXCLASSFit} and \texttt{myXCLASSMapFit} (\texttt{myXCLASSMapRedoFit}) functions and we are exploring the use of graphics processing units (GPUs).\\

\noindent Generally, \texttt{MAGIX} is controlled by different XML files: The so-called instance file includes the names, initial values (and ranges) for all model parameters. Additionally, the instance file indicates the parameters which should be optimized by \texttt{MAGIX}. In addition the experimental XML file containing settings for the import of observational data, that is, path(s) and name(s) of the data file(s), format(s), range(s) etc. Furthermore, the algorithm control file, defining the algorithm or algorithm sequence which should be used for the optimization together with the settings for each algorithm.

The new \texttt{MAGIX} function offers the possibility to use \texttt{MAGIX} within CASA with a previously registered external model program. For that purpose, the user has to provide the XML files, described above, and passes their path and file names to the \texttt{MAGIX} function, see \ref{app:magix}.

In general, XCLASS creates job directories for some functions (\texttt{MAGIX}, \texttt{myXCLASSFit}, and \texttt{myXCLASSMapFit}) where all log and output files are stored. The job directories are created in the XCLASS run directory defined by the environment variable \texttt{myXCLASSRunDirectory}. The name of a job directory contains the date and time of the function execution followed by a unique job number.

    \section{Other functions}\label{sec:NewFunc}

The {\sc XCLASS} interface for CASA includes two new functions (\texttt{myXCLASSFit} and \texttt{myXCLASSMapFit}) which provide a simplified interface for MAGIX using the myXCLASS program.

The \texttt{myXCLASSFit} function offers the possibility to fit multiple ranges in multiple files from multiple telescopes simultaneously using \texttt{MAGIX} in conjunction with the \texttt{myXCLASS} program. It starts \texttt{MAGIX} using the Levenberg-Marquardt algorithm to optimize the input parameters defined in an extended molfit file to achieve a good description of the observational data. The user has to specify the maximum number of iterations, the path and file name of the observational data file (or of a \texttt{MAGIX} experimental XML file, see Sect.~\ref{sec:MAGIX}), and the path and name of an extended molfit file. In contrast to the molfit file described in Sect.~\ref{myxclass:molfit}, the extended molfit file, see appendix~\ref{app:myxclassfit}, contains one additional column on the left side of each parameter (except for the core/foreground flag) of each component. The additional column defines the range for each parameter. The definition of ranges guarantees that the fits does not run out of the defined parameter ranges. The limits, that is, the lowest and highest allowed values, for the parameters source size $\theta_{m, c}$, column density $N_{\rm tot}$, and hydrogen column density $N_H$ are determined by
\begin{align}
  {\rm lower \, limit} &= |{\rm parameter \, value}| \, / \, ({\rm range \,value}) \nonumber\\
  {\rm upper \, limit} &= |{\rm parameter \, value}| \, \times \, ({\rm range \, value}) \nonumber.
\end{align}
The limits for the other parameters are calculated by simply adding or subtracting the value of the additional column to or from the value itself. All calculated lower limits which are less than zero are set to zero (except for velocity offset). If the range of a certain parameter is set to zero, the corresponding parameter is kept constant and is not optimized by the \texttt{myXCLASSFit} function. Additionally, the \texttt{myXCLASSFit} function offers the possibility to fit the iso ratios in the same fit process as well. For that purpose, the user has to add two additional columns in the iso ratio file on the right side of the third column. The fourth and fifth column defines the lower and upper limit for each iso ratio, respectively. If the lower limit is equal to the upper limit or if the lower limit is higher than the upper limit, the corresponding iso ratio is not optimized by the \texttt{myXCLASSFit} function. In general, the fit procedure stops, if the maximum number of iterations is reached, or if $\chi^2$ drops below $10^{-7}$ (or an user defined value). Additionally, the \texttt{myXCLASSFit} function offers the possibility to use another algorithm or an algorithm chain or tree by defining the path and name of a \texttt{MAGIX} algorithm XML file, see Sect.~\ref{sec:MAGIX}. Furthermore, the function offers the possibility to fit multiple frequency ranges simultaneously by using a \texttt{MAGIX} experimental XML file, see Sect.~\ref{sec:MAGIX}. Finally, the \texttt{myXCLASSFit} function returns the optimized molfit file from the best fit and the corresponding modeled spectra as python arrays. (Here, the phrase ``best fit'' is connected to the application of an algorithm chain or tree: Using an algorithm chain or tree offers the possibility to use more than one algorithm to fit the data. Each application of an algorithm is connected with a fit represented by a certain $\chi^2$ value. The \texttt{myXCLASSFit} function reads in the $\chi^2$ values of all optimization loops of all applied algorithm and determines the lowest $\chi^2$ value. The result of the ``best fit'' is than given by the parameter vector which belongs to the lowest $\chi^2$ value.)\\

    \begin{figure}[t]
       \centering
       \includegraphics[width=0.51\textwidth]{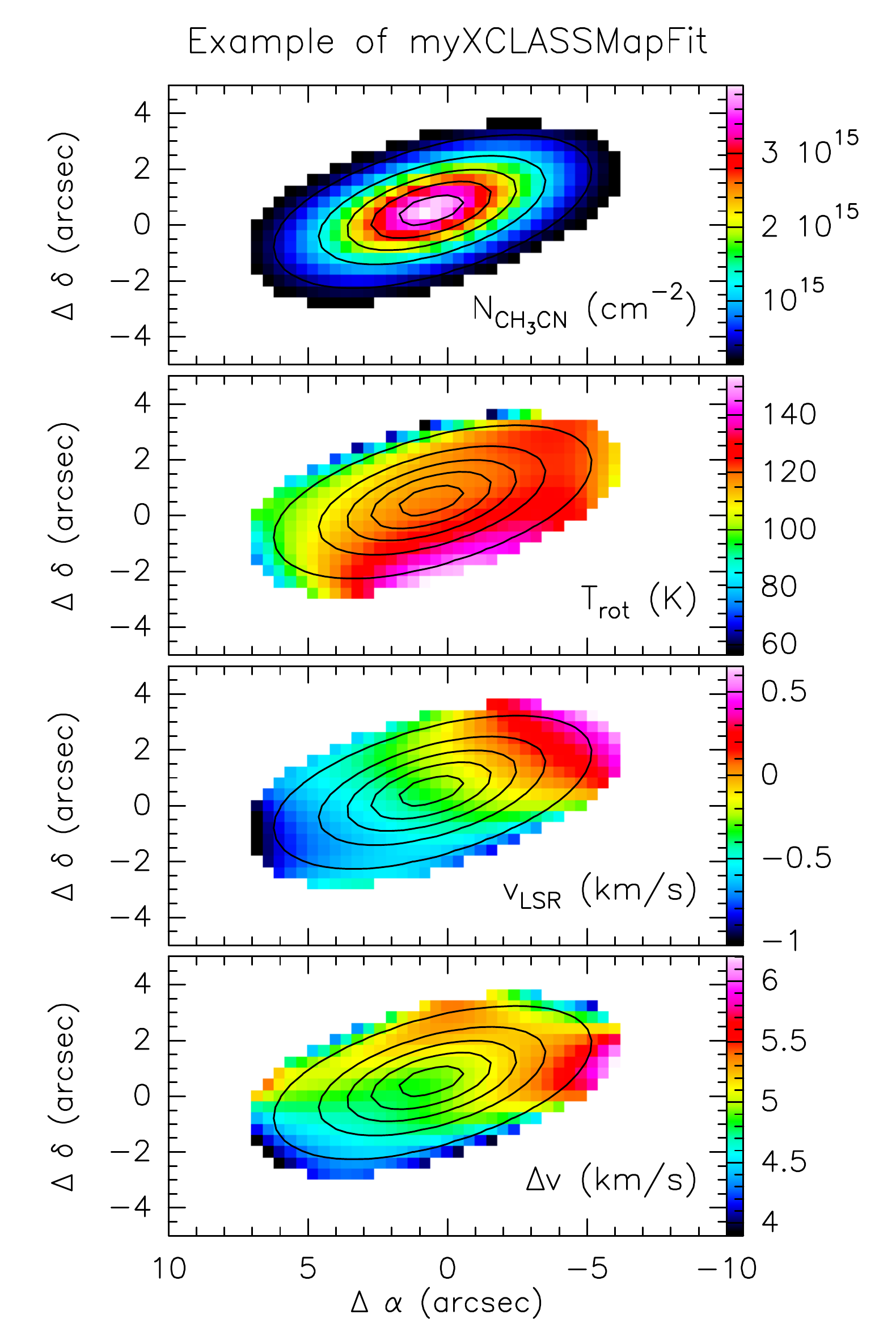}\\
       \caption{Example of parameter maps created by the \texttt{myXCLASSMapFit} function. Here, the K-ladder structure of the CH$_3$CN(12-11) transition (plus isotopologues) in G75.78+0.34 was fitted at around 220~GHz (observed with the SMA). The temperature gradient is expected as also seen in different NH$_3$ lines \cite{SanchezMonge1, SanchezMonge2}.}
       \label{fig:myxclassmapfit}
    \end{figure}

In addition to the \texttt{myXCLASSFit} function which is useful to fit single spectra, the {\sc XCLASS} interface for CASA contains the \texttt{myXCLASSMapFit} function which fits one or more complete (FITS) data cubes. Here, the \texttt{myXCLASSMapFit} function reads in the data cube(s), extracts the spectra for each pixel and fits these spectra separately using the Levenberg-Marquardt (or another) algorithm (chain or tree). The optimization procedure for each pixel stops, if the maximum number of iterations is reached, or if the $\chi^2$ value drops below $10^{-7}$ (or below an user defined value). In analogous to the \texttt{myXCLASSFit} function the fit parameters and their ranges for the \texttt{myXCLASSMapFit} function are defined in an extended molfit (and iso ratio) file as well. Additionally, the \texttt{myXCLASSMapFit} function offers the possibility to limit the fit to certain frequency ranges of a spectrum and to an user defined region of the cube(s). In order to reduce the computation effort, the user can exclude pixels by defining a threshold for the min. intensity of a pixel, that is, if the maximum intensity of a pixel is below the user defined threshold, the corresponding pixel is not fitted by the the \texttt{myXCLASSMapFit} function. In general, the \texttt{myXCLASSMapFit} function assumes, that the first three axes of the data cube(s) describe the right ascension, the declination, and the frequency, respectively. If the frequencies are not given in MHz, the FITS header has to contain the definition of the CUNIT3 command which defines the unit of the frequency axis. If more than one data cube is specified in the XML file, the data cubes must describe the same map, that is, covered spatial area and the resolution have to be identical. The different data cubes are allowed to differ only in the frequency axis. At the end of the whole fit procedure, the \texttt{myXCLASSMapFit} function creates FITS images for each free parameter of the best fit, where each pixel corresponds to the value of the optimized parameter taken from the best fit for this pixel, see Fig.~\ref{fig:myxclassmapfit}. Furthermore, the \texttt{myXCLASSMapFit} function creates FITS cubes for each fitted data cube, where each pixel contains the modeled spectrum. Finally, the \texttt{myXCLASSMapFit} function creates one FITS image which describes the quality of the fit for each pixel. Here, each pixel corresponds to the $\chi^2$ value of the best fit for this pixel. Applications of this are temperature maps, but also first and second moment maps, which are based on the simultaneous fitting of many lines, and are fairly robust against line confusion and blending of single lines.\\

In order to improve the results of a previous application of the \texttt{myXCLASSMapFit} function the {\sc XCLASS} interface contains the \texttt{myXCLASSMapRedoFit} function which offers the possibility to redo one or more so-called pixel fits of a previous myXCLASSMapFit run. The function performs fits for the selected pixels and recreates the different parameter maps using the new optimized parameter values.

    \section{Example application}


    \begin{figure}[t]
      \centering
      \includegraphics[width=0.44\textwidth]{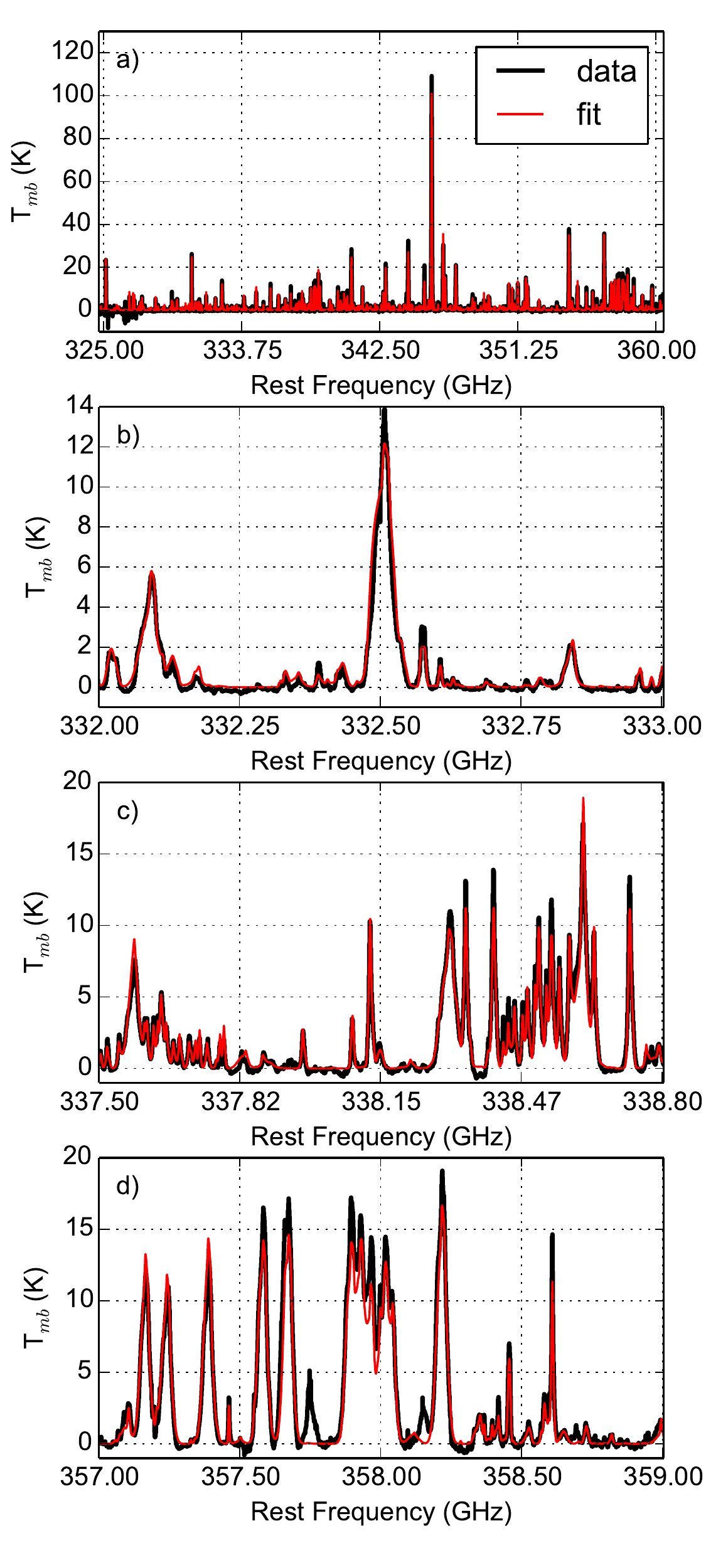}\\
      \caption{Result of the spectral line survey of the high-mass star-forming region Orion KL between 325 to 360 GHz using the \texttt{myXCLASSFit} function. Here, we used 24 molecules in conjunction with 245 isotopologues. In panel~a) the whole spectral line survey (back) is shown together with the result of \texttt{myXCLASSFit} function (red). Panel~b),~c) and~d) show fit examples around 332.5~GHz, 338.15~GHz, and 358.0~GHz, respectively.}
      \label{fig:Examples}
    \end{figure}

The {\sc XCLASS} interface for CASA was used to redo the analysis of the spectral line survey of the high-mass star-forming region Orion~KL from 325 to 360~GHz as reported by \citet{Schilke1997}, which was done without using the myXCLASS program nor optimization packages like MAGIX.

In our fit, shown in Fig.~\ref{fig:Examples}, we used the following 25 molecules: C$_2$H$_3$CN, C$_2$H$_5$CN, CCH, CH$_3$CCH, CH$_3$CN, CH$_3$OCH$_3$, CH$_3$OCHO, CH$_3$OH, CN, CO, CS, H$_2$CO, H$_2$CS, H$_2$O, HCCCN, HCN, HCO$^+$, HCS$^+$, HDO, HNCO, NO, OCS, SiO, SO$_2$, and SO. The different molecules were described by 44 components in total, see Tab~\ref{tab:MolfitParam}. Additionally, we have taken 241 isotopologues and higher vibrated states into account where we assumed the following ratios: [$^{12}$C]/[$^{13}$C]~=~45 \cite{Crockett2014a}, [$^{16}$O]/[$^{18}$O]~=~250 \cite{Crockett2014a}, [$^{16}$O]/[$^{17}$O]~=~2625 \cite{Tercero2010}, [$^{32}$S]/[$^{33}$S]~=~75 \cite{Crockett2014a}, [$^{32}$S]/[$^{34}$S]~=~22.5 \cite{Schilke1997}, [$^{14}$N]/[$^{15}$N]~=~234 \cite{Crockett2014a}, [$^{28}$Si]/[$^{29}$Si]~=~20 \cite{Schilke1997}, and [$^{28}$Si]/[$^{30}$Si]~=~30 \cite{Anders1989}. In order to reduce the computational effort, we assumed a ratio of one for all vibrational excited transitions, that is, we used the same column densities as for the corresponding molecules.\\

%
Additionally, we were able to (partly) identify 35 of the 57 unidentified lines (U-lines) reported by \citet{Schilke1997}, see Tab.~\ref{tab:ULines}.  Here, we marked an unidentified line as (partly) identified, if the modeled integrated intensity of a component $\left(\int {\rm T}_R^{\rm fit} \, d {\rm v}\right)$ covers at least 10 \% of the integrated line intensity $\left(\int {\rm T}_R^{\rm Schilke} \, d {\rm v}\right)$. We note that, the majority of the unidentified lines contain contributions from more than one component. Contributions of more than 100 \% are caused by inaccuracies in the fit. For 19 U-lines the modeled spectrum describes less than 50 \% of the integrated line intensity. Nevertheless, we are able to (partly) describe all unidentified lines, except the line at 348358~MHz, with an integrated line intensity of more than 10~K~km~s$^{-1}$. The improved identification may caused by an updated and enhanced line catalog.\\

%
In the following we will compare our fit results, given in Tab.~\ref{tab:MolfitParam}, with those  reported by \citet{Schilke1997}. Our calculated errors describe the 2$\sigma$ credibility interval for each parameter, respectively. Large asymmetric error ranges for some column densities (e.g.,\ SO) are caused by the fact that some lines start to become optically thick which prevents an accurate parameter estimation. For these lines we can only determine lower limits of the corresponding column densities.
\begin{table*}[!p]
\centering
\caption{\label{tab:MolfitParam} Fit results of the molfit file parameters used for the Orion-KL model. Here, we describe only those molecules, which show at least three transitions within the covered frequency range.}
\begin{small}
\begin{tabular}{lccccc}
\hline
\hline
Molecule      & $\theta^{m,c}$ (arcsec) & T$_{\rm ex}^{m,c}$ (K) &   N$_{\rm tot}^{m,c}$ (cm$^{-2})$ & $\Delta$ v$^{m,c}$ (km s$^{-1}$) & v$_{\rm LSR}^{m,c}$ (km s$^{-1}$)\\
\hline
CH$_3$CN      &      3$^{+37.6}_{-2.0}$    &  245$^{+24.1}_{-18.2}$         &  3.1 $(14) \, ^{+1.2 (17)}_{-1.0 (14)}$  &   10$^{+4.0}_{-2.2}$  &    7$^{+1.8}_{-1.3}$\\
              &                    -       & [445$^{+36}_{-36}$]            &  [1.7$(15) \, ^{+0.7 (15)}_{-0.7 (15)}$] &                     - & - \\
CH$_3$CCH     &     70$^{+9.8}_{-69.3}$    &   21$^{+26.3}_{-14.5}$         &  4.2 $(16) \, ^{+8.6 (17)}_{-6.4 (15)}$  &    6$^{+2.0}_{-6.1}$  &   10$^{+1.4}_{-1.0}$\\
              &                      -     &  [65$^{+9}_{-9}$]              &  [4.3$(15) \, ^{+2.2 (15)}_{-2.2 (15)}$] &                     - & - \\
HCCCN         &     81$^{+37.0}_{-13.6}$   &   72$^{+24.2}_{-15.1}$         &  1.0 $(15) \, ^{+7.8 (16)}_{-2.5 (14)}$  &   11$^{+3.5}_{-3.1}$  &    5$^{+1.9}_{-1.4}$\\
              &      2$^{+35.8}_{-1.3}$    &  330$^{+20.0}_{-7.1}$          &  5.9 $(14) \, ^{+4.8 (16)}_{-1.5 (14)}$  &   21$^{+4.1}_{-2.9}$  &    4$^{+1.7}_{-1.4}$\\
              &                    -       & [225$^{+200}_{-200}$]          &  [5.2$(15) \, ^{+7.8 (15)}_{-7.8 (15)}$] &                     - & - \\
CCH           &    371$^{+29.0}_{-326.1}$  &   31$^{+265.5}_{-29.7}$        &  1.6 $(11) \, ^{+1.6 (11)}_{-2.7 (8)}$   &    3$^{+4.3}_{-3.0}$  &   10$^{+11.0}_{-10.3}$\\
              &    326$^{+63.7}_{-296.0}$  &  330$^{+20.0}_{-6.5}$          &  7.9 $(14) \, ^{+3.3 (16)}_{-1.7 (14)}$  &    3$^{+2.4}_{-2.7}$  &   10$^{+2.4}_{-1.9}$\\
OCS           &    180$^{+8.4}_{-179.0}$   &   65$^{+25.8}_{-13.7}$         &  1.8 $(16) \, ^{+3.1 (18)}_{-5.2 (15)}$  &   16$^{+3.4}_{-3.2}$  &    7$^{+1.6}_{-1.3}$\\
              &    261$^{+125.2}_{-260.0}$ &  316$^{+1.0}_{-314.9}$         &  1.1 $(15) \, ^{+4.9 (16)}_{-2.2 (14)}$  &    5$^{+3.5}_{-3.4}$  &    8$^{+2.2}_{-1.7}$\\
              &                       -    &  [83$^{+30}_{-30}$]            &  [1.8$(16) \, ^{+1.8 (16)}_{-1.8 (16)}$] &                     - & - \\
HNCO          &    332$^{+57.7}_{-331.0}$  &  160$^{+25.8}_{-20.7}$         &  4.7 $(14) \, ^{+5.2 (16)}_{-1.3 (14)}$  &    6$^{+3.6}_{-3.3}$  &    7$^{+2.0}_{-1.3}$\\
              &     20$^{+29.8}_{-16.7}$   &   16$^{+25.8}_{-14.9}$         &  7.4 $(16) \, ^{+1.3 (18)}_{-1.1 (16)}$  &   16$^{+3.4}_{-3.3}$  &    7$^{+1.8}_{-1.4}$\\
              &                       -    & [160$^{+14}_{-14}$]            &  [1.3$(15) \, ^{+2.3 (15)}_{-2.3 (15)}$] &                     - & - \\
              &                       -    &  [36$^{+5}_{-5}$]              &  [5.7$(15) \, ^{+3.5 (15)}_{-3.5 (15)}$] &                     - & - \\
              &                       -    & [277$^{+100}_{-100}$]          &  [1.3$(15) \, ^{+0.8 (15)}_{-0.8 (15)}$] &                     - & - \\
              &                       -    & [190$^{+40}_{-40}$]            &  [1.2$(15) \, ^{+0.6 (15)}_{-0.6 (15)}$] &                     - & - \\
CN            &     42$^{+27.9}_{-22.3}$   &   43$^{+21.9}_{-18.1}$         &  1.7 $(14) \, ^{+6.2 (15)}_{-3.7 (13)}$  &    4$^{+4.0}_{-2.3}$  &   10$^{+1.7}_{-2.1}$\\
              &     10$^{+30.4}_{-9.5}$    &   15$^{+22.6}_{-14.2}$         &  1.3 $(15) \, ^{+2.7 (16)}_{-2.2 (14)}$  &   22$^{+3.7}_{-3.5}$  &    8$^{+2.7}_{-1.1}$\\
H$_2$CO       &    249$^{+2.2}_{-248.0}$   &  150$^{+24.9}_{-24.3}$         &  8.4 $(14) \, ^{+8.4 (17)}_{-3.3 (14)}$  &   24$^{+3.8}_{-3.5}$  &    8$^{+1.9}_{-1.3}$\\
              &    275$^{+112.0}_{-274.0}$ &  130$^{+23.1}_{-24.6}$         &  1.8 $(14) \, ^{+1.4 (16)}_{-5.1 (13)}$  &    7$^{+3.5}_{-3.5}$  &    9$^{+2.0}_{-1.3}$\\
              &                       -    & [H$_2^{13}$CO: 230$^{+105}_{-105}$]  & [H$_2^{13}$CO: 6.3 $(14) \, ^{+4.7 (14)}_{-4.7 (14)}$]     & - & - \\
HDO           &    150$^{+101.6}_{-75.4}$  &   11$^{+171.1}_{-9.6}$         &  5.0 $(10) \, ^{+4.9 (17)}_{-5.0 (10)}$  &    4$^{+17.0}_{-4.4}$ &   18$^{+1.6}_{-0.7}$\\
              &     47$^{+12.4}_{-45.5}$   &  270$^{+17.3}_{-268.9}$        &  5.7 $(15) \, ^{+6.5 (17)}_{-1.4 (15)}$  &    9$^{+4.9}_{-2.7}$  &    7$^{+1.4}_{-1.5}$\\
              &                       -    & [272$^{+200}_{-200}$]          &  [0.9$(16) \, ^{+1.3 (16)}_{-1.3 (16)}$] &                     - & - \\
H$_2$CS       &    105$^{+29.2}_{-19.3}$   &  105$^{+26.7}_{-14.7}$         &  2.1 $(14) \, ^{+4.4 (16)}_{-7.1 (13)}$  &    6$^{+3.5}_{-3.4}$  &    8$^{+1.9}_{-1.3}$\\
              &                       -    &  [93$^{+19}_{-19}$]            &  [8.5$(14) \, ^{+3.5 (14)}_{-3.5 (14)}$] &                     - & - \\
NO            &     38$^{+299.8}_{-36.9}$  &   47$^{+49.2}_{-15.0}$         &  3.3 $(12) \, ^{+6.6 (14)}_{-3.3 (12)}$  &   10$^{+1.2}_{-10.4}$ &    5$^{+3.2}_{-0.2}$\\
              &     29$^{+26.5}_{-24.4}$   &  196$^{+23.9}_{-21.5}$         &  1.2 $(17) \, ^{+1.3 (19)}_{-3.3 (16)}$  &   22$^{+3.7}_{-3.4}$  &    8$^{+1.3}_{-0.9}$\\
SO            &     10$^{+28.0}_{-9.5}$    &   65$^{+24.4}_{-14.7}$         &  7.1 $(16) \, ^{+9.3 (19)}_{-3.1 (16)}$  &   25$^{+3.9}_{-2.1}$  &    9$^{+3.0}_{-0.9}$\\
              &      5$^{+30.2}_{-4.0}$    &  264$^{+20.5}_{-20.5}$         &  2.9 $(16) \, ^{+3.2 (19)}_{-1.2 (16)}$  &   15$^{+3.5}_{-2.4}$  &    9$^{+1.8}_{-1.9}$\\
              &                       -    &  [27$^{+1}_{-1}$]              &  [6.3$(16) \, ^{+0.7 (16)}_{-0.7 (16)}$] &                     - & - \\
CH$_3$OCHO    &      3$^{+38.8}_{-2.3}$    &  117$^{+24.0}_{-20.6}$         &  9.3 $(15) \, ^{+6.6 (18)}_{-3.4 (15)}$  &    5$^{+3.9}_{-2.4}$  &    8$^{+1.0}_{-1.1}$\\
              &                       -    &  [98$^{+3}_{-3}$]              &  [1.5$(16) \, ^{+0.1 (16)}_{-0.1 (16)}$] &                     - & - \\
CH$_3$OCH$_3$ &      3$^{+40.0}_{-2.0}$    &  136$^{+24.6}_{-20.4}$         &  1.0 $(16) \, ^{+1.9 (17)}_{-1.8 (15)}$  &    3$^{+3.9}_{-2.5}$  &    9$^{+1.4}_{-0.9}$\\
              &     24$^{+25.9}_{-23.2}$   &   24$^{+22.5}_{-22.3}$         &  2.8 $(15) \, ^{+10.0 (16)}_{-4.9 (14)}$ &    6$^{+3.9}_{-2.5}$  &    9$^{+1.3}_{-1.0}$\\
              &                       -    &  [89$^{+5}_{-5}$]              &  [1.8$(16) \, ^{+0.2 (16)}_{-0.2 (16)}$] &                     - & - \\
C$_2$H$_5$CN  &     98$^{+15.0}_{-35.5}$   &  127$^{+23.3}_{-21.6}$         &  9.9 $(14) \, ^{+1.1 (18)}_{-3.9 (14)}$  &   13$^{+3.5}_{-3.1}$  &    5$^{+1.7}_{-1.4}$\\
              &                       -    &  [99$^{+3}_{-3}$]              &  [1.3$(16) \, ^{+0.2 (16)}_{-0.2 (16)}$] &                     - & - \\
C$_2$H$_3$CN  &     15$^{+29.1}_{-13.6}$   &  174$^{+31.4}_{-16.1}$         &  8.7 $(13) \, ^{+3.4 (15)}_{-1.9 (13)}$  &    4$^{+1.0}_{-0.7}$  &    6$^{+1.0}_{-0.9}$\\
              &     19$^{+29.2}_{-17.2}$   &  126$^{+23.3}_{-21.5}$         &  3.5 $(14) \, ^{+1.6 (16)}_{-7.6 (13)}$  &   12$^{+1.0}_{-0.7}$  &    3$^{+0.9}_{-1.2}$\\
              &                       -    &  [96$^{+5}_{-5}$]              &  [8.2$(14) \, ^{+1.5 (14)}_{-1.5 (14)}$] &                     - & - \\
CH$_3$OH      &     12$^{+27.6}_{-10.5}$   &   25$^{+25.8}_{-14.4}$         &  9.2 $(15) \, ^{+7.3 (17)}_{-2.8 (15)}$  &    5$^{+3.7}_{-2.6}$  &    9$^{+1.8}_{-1.4}$\\
              &      3$^{+37.1}_{-2.1}$    &  189$^{+26.0}_{-20.0}$         &  4.5 $(16) \, ^{+1.6 (20)}_{-2.1 (16)}$  &    8$^{+3.8}_{-2.5}$  &    7$^{+1.5}_{-1.4}$\\
              &      4$^{+36.9}_{-2.5}$    &  161$^{+28.1}_{-18.3}$         &  1.3 $(16) \, ^{+1.3 (18)}_{-4.2 (15)}$  &    2$^{+3.9}_{-2.4}$  &    9$^{+2.1}_{-1.8}$\\
              &                       -    & [188$^{+3}_{-3}$]              &  [7.0$(16) \, ^{+0.2 (16)}_{-0.2 (16)}$] &                     - & - \\
SO$_2$        &      3$^{+40.6}_{-1.9}$    &  140$^{+29.7}_{-16.5}$         &  2.3 $(16) \, ^{+1.3 (19)}_{-9.3 (15)}$  &    6$^{+3.9}_{-2.4}$  &    5$^{+1.0}_{-1.1}$\\
              &      9$^{+29.3}_{-7.8}$    &  173$^{+29.9}_{-18.9}$         &  4.1 $(16) \, ^{+3.5 (21)}_{-2.1 (16)}$  &   31$^{+4.6}_{-2.8}$  &   11$^{+0.8}_{-1.1}$\\
              &     16$^{+29.4}_{-14.6}$   &   23$^{+25.4}_{-14.9}$         &  5.4 $(16) \, ^{+3.9 (20)}_{-2.7 (16)}$  &   21$^{+4.9}_{-2.5}$  &   11$^{+0.8}_{-1.1}$\\
              &                       -    & [124$^{+3}_{-3}$]              &  [7.7$(16) \, ^{+0.5 (16)}_{-0.5 (16)}$] &                     - & - \\
\hline
\end{tabular}
\end{small}
\tablefoot{The column densities are given on a main beam brightness line temperature scale, i.e.,\ each column density (exponents are described by round brackets, e.g.,\ $1.4(16) = 1.4 \times 10^{16}$) is multiplied with the beam filling factor of the corresponding component, to make them comparable to the results given by \citet{Schilke1997}, which were indicated here within squared brackets. Additionally, the errors of the fit parameters (indicated by the sub-(left) and superscript (right) values) were determined using the error estimation algorithm included in the MAGIX package.}
\end{table*}

\begin{table}[!p]
\centering
\caption{\label{tab:ULines} New identified lines. Here, an unidentified line is marked as identified, if the modeled integrated intensity of at least one molecule covers at least 10 \%. The numbers in squared brackets describe the components of the corresponding molecules, respectively.}
\begin{small}
\begin{tabular}{cccc}
\hline
\hline
$\nu$  & $\int$ T$_R^{\rm Schilke} d$v & $\int$ T$_R^{\rm fit} d$v & Molecule\\
(MHz)  & (K km s$^{-1}$)               & (K km s$^{-1}$)           &\\
\hline
330715 &  1.7 & 3.3 (195.7 \%) & CH$_3$OCHO,v$_{18}$=1, [1]\\
332789 &  3.8 & 1.0 (25.9 \%) & C$_2$H$_3$CN, [2]\\
333865 &  8.9 & 1.4 (15.5 \%) & $^{34}$SO, [1]\\
334140 &  1.6 & 1.5 (91.9 \%) & CH$_3$OCHO,v$_{18}$=1, [1]\\
335335 &  6.2 & 0.8 (12.7 \%) & C$_2$H$_5$CN, [1]\\
335703 &  4.6 & 1.0 (22.2 \%) & CH$_3$OH, [2]\\
335840 &  2.9 & 0.8 (27.3 \%) & CH$_3$OCHO, [1]\\
337744 &  4.2 & 1.3 (31.6 \%) & CH$_3$OH,v$_{12}$=1, [2]\\
337839 &  3.3 & 4.9 (149.7 \%) & HCCCN,v$_{7}$=1, [2]\\
       &      & 3.1 (94.2 \%) & CH$_3$OH, [2]\\
       &      & 0.8 (23.0 \%) & CH$_3$OH, [3]\\
339138 &  3.3 & 0.5 (14.0 \%) & $^{13}$CH$_3$CN-A [1]\\
339527 &  8.6 & 1.2 (13.4 \%) & CN, [2]\\
340496 & 15.9 & 6.9 (43.5 \%) & C$_2$H$_5$CN, [1]\\
340872 &  4.6 & 0.7 (14.4 \%) & $^{33}$SO, [1]\\
341472 & 10.9 & 10.3 (94.9 \%) & C$_2$H$_5$CN, [1]\\
341482 & 11.7 & 2.1 (18.1 \%) & C$_2$H$_5$CN, [1]\\
341499 &  0.4 & 0.2 (54.1 \%) & CH$_3$OCHO,v$_{18}$=1, [1]\\
342129 &  4.1 & 0.6 (14.7 \%) & C$_2$H$_3$CN, [2]\\
       &      & 0.7 (16.2 \%) & CH$_3$OCHO,v$_{18}$=1, [1]\\
       &      & 0.4 (10.3 \%) & C$_2$H$_3$CN, [1]\\
342290 &  3.9 & 3.7 (96.1 \%) & CH$_3$OCHO,v$_{18}$=1, [1]\\
342486 &  4.7 & 0.9 (20.0 \%) & CH$_3$OCHO,v$_{18}$=1, [1]\\
343202 & 21.8 & 13.1 (60.2 \%) & C$_2$H$_5$CN, [1]\\
343665 &  7.8 & 1.7 (21.4 \%) & CH$_3$OCHO,v$_{18}$=1, [1]\\
344773 &  ... & 0.2           & $^{34}$SO$_2$, [2]\\
344788 &  ... & 1.2           & $^{34}$SO$_2$, [2]\\
       &      & 1.0           & CH$_3$OCHO, [1]\\
       &      & 0.2           & $^{34}$SO$_2$, [3]\\
344796 &  ... & 2.5           & $^{34}$SO$_2$, [2]\\
       &      & 0.8           & $^{34}$SO$_2$, [3]\\
347446 &  4.7 & 2.2 (45.9 \%) & C$_2$H$_3$CN, [2]\\
       &      & 0.8 (17.8 \%) & CH$_3$OH,v$_{12}$=1, [2]\\
       &      & 0.7 (15.0 \%) & C$_2$H$_5$CN, [1]\\
348084 &  2.4 & 0.5 (22.7 \%) & $^{34}$SO$_2$, [2]\\
       &      & 2.3 (95.5 \%) & CH$_3$OCHO,v$_{18}$=1, [1]\\
348373 &  ... & 31.0           & SO$_2$, [2]\\
       &      & 0.2           & SO$^{18}$O, [3]\\
       &      & 0.2           & HCCCN,v$_{7}$=3, [2]\\
       &      & 0.2           & SO$^{18}$O, [2]\\
       &      & 0.4           & SO$^{18}$O, [1]\\
       &      & 0.2           & CCH,v$_{2}$=2, [2]\\
       &      & 0.3           & C$_2$H$_5$CN, [1]\\
350170 &  2.8 & 2.3 (81.3 \%) & CH$_3$CN,v$_{8}$=1, [1]\\
       &      & 1.5 (53.3 \%) & C$_2$H$_5$CN, [1]\\
351490 &  7.2 & 1.1 (15.8 \%) & $^{33}$SO$_2$, [3]\\
351540 & 10.8 & 5.0 (45.9 \%) & HNCO, [1]\\
       &      & 3.1 (28.6 \%) & $^{33}$SO$_2$, [3]\\
       &      & 1.3 (11.9 \%) & C$_2$H$_5$CN, [1]\\
353166 &  5.0 & 1.2 (23.3 \%) & C$_2$H$_3$CN, [2]\\
354129 &  7.8 & 5.6 (72.3 \%) & CH$_3$OH,v$_{12}$=1, [2]\\
       &      & 1.2 (14.7 \%) & CH$_3$OH,v$_{12}$=1, [3]\\
355851 &  3.2 & 0.4 (12.8 \%) & OC$^{34}$S, [1]\\
356644 &  5.3 & 1.1 (21.4 \%) & HCCCNv$_6$=1, v$_7$=1, [2]\\
358356 & 14.1 & 1.6 (11.1 \%) & $^{34}$SO$_2$, [2]\\
       &      & 1.9 (13.5 \%) & $^{34}$SO$_2$, [1]\\
       &      & 3.3 (23.5 \%) & C$_2$H$_5$CN, [1]\\
\hline
\end{tabular}
\end{small}
\tablefoot{Due to inaccuracies in the fit we overestimated the integrated line intensities of some weak lines.
}
\end{table}

%
In contrast to \citet{Schilke1997} we were not able to identify NH$_2$CHO,  C$_2$H$_5$OH, and HCOOH because of their weak lines which do not allow a distinct estimation of their contribution to the given spectrum. But we took CCH, CN, and NO into account.

%
Our derived excitation temperatures and beam averaged column densities for H$_2$CS and CH$_3$OCHO agree (within the given error ranges) with Schilke's result.
%
%
In order to achieve a good description for HCCCN, HDO, and OCS we used two components, respectively. For HCCCN and HDO we find that the second components show similar temperatures and column densities compared to Schilke's results. Due to the large error ranges for OCS we are not able to identify the corresponding component for Schilke's result although the temperatures as well as column densities are comparable within the given error ranges.

%
Additionally, we used H$_2$CO with two components instead of H$_2^{13}$CO with one component. Using the aforementioned isotopologue ratio of [$^{12}$C]/[$^{13}$C]~=~45 we find that the temperature and column density for the first component of H$_2$CO agree with those described by \citet{Schilke1997}.

%
As shown in Tab.~\ref{tab:MolfitParam}, we used three components to describe the contribution of CH$_3$OH and SO$_2$. Schilke's result for CH$_3$OH corresponds to our second component whereas the excitation temperature and beam averaged column density for the first component of SO$_2$ agree with those reported by \citet{Schilke1997}. Furthermore, the described temperatures and column densities (after scaling with the isotopologue ratios, mentioned above) for $^{34}$SO$_2$ match our results nicely. For $^{33}$SO$_2$ the derived excitation temperature do not coincide with Schilke's result.

%
In order to model the contribution of HNCO we used only two instead of four components. We see that the temperatures and column densities of Schilke's first, third, and fourth component match our results for the first component. But, the column density of Schilke's second component clearly does not agree with our column density for the second component although the excitation temperatures are comparable.

%
Although we find similar results for the majority of molecules we derived different excitation temperatures for CH$_3$CN, CH$_3$CCH, SO, CH$_3$OCH$_3$, C$_2$H$_3$CN, and C$_2$H$_5$CN. In contrast to CH$_3$CCH, SO, C$_2$H$_3$CN, C$_2$H$_5$CN, and CH$_3$OCH$_3$ where we find discrepancies with less than 24~K we find a striking difference for CH$_3$CN. Already \cite{Schilke1999} noticed that the value of 445~K based on rotation diagram analysis was wrong due to a neglect of the line opacity. They presented a corrected fit using the opacities derived from CH$_3$CN/CH$_3^{13}$CN, yielding 160~K, which differs from the present value because (i) a [$^{12}$C]/[$^{13}$C] ratio of 60 (instead of 45) was used then - which increases the correction due to opacity, and (ii) the correction was done by hand, not taking the individual lines as well into account as the present study. Not taking into account the opacities also underestimates the column density, by an order of magnitude in the present case. Additionally, we derived a column density for CH$_3$CCH which is nearly an order of magnitude higher than Schilke's result.\\

%
For optically thin, unblended lines (like for H$_2$S and CH$_3$OCHO), the rotation diagram method gives the same results as the XCLASS fit, but requires more effort through manual fitting of all lines. High opacities, while they can be corrected in rotation diagrams \cite{Schilke1997, GoldsmithLanger1999}, this is very cumbersome to do by hand, and is much more robustly achieved through the use of XCLASS - in this data set, CH$_3$CN and CH$_3$OH are prominent examples.  Lastly, the identification of species with weak features is only possible reliably in line-rich spectra if all the other species are modeled as well, as can be seen by the XCLASS fits reported in \cite{Belloche2008, Belloche2009, Belloche2013}. All this together with the capability of producing credibility intervals demonstrates that the methods of XCLASS give much more accurate and reliable results than those achieved with rotational diagram methods.\\


    \section{Conclusions}

We presented the {\sc XCLASS} interface for CASA which provides powerful new functions for the CASA distribution for analyzing spectral surveys. The toolbox includes the \texttt{myXCLASS} program which models observational data by solving the radiative transfer equation for an isothermal object in one dimension (detection equation). The \texttt{myXCLASS} program is able to model a huge number of molecules simultaneously, whereat the contribution of each molecule can be described by an arbitrary number of components. These can usually be distinguished by different radial velocities and do not interact with each other radiatively but superimpose in the model. The \texttt{myXCLASS} program depends on a number of parameters, which are partially taken from an embedded SQLite3 database containing entries from CDMS and JPL through the VAMDC portal. The other input parameters are read from an user defined ASCII file containing parameters for each molecule and component. In order to achieve a reasonable description of observational data one has to partially optimize the user defined parameters. For that purpose the toolbox contains an interface for the model optimizer package \texttt{MAGIX}, which helps to find the best description of the data using a certain model, that is, finding the parameter set that most closely reproduces the data. Therefore, the toolbox provides two functions to model one or more single spectra or data cubes simultaneously using the \texttt{myXCLASS} program in conjunction with one of the optimization algorithms (or combinations of them in an algorithm chain or tree) included in the \texttt{MAGIX} package. The combination of \texttt{myXCLASS} and \texttt{MAGIX} prepare the way for an automatic line identification routine which is essential for a spectral survey on large spectral cubes produced by ALMA. Additionally, the \texttt{MAGIX} package offers the possibility to apply more elaborate programs (such as LIME) to model astronomical data as well.


Enhancements are being worked on a 3-d version for defining sub-beam structure. Additionally, we work on an automatic line identification function for XCLASS, which will be described in a subsequent paper.

    \begin{acknowledgements}
          The authors would like to thank Anika Schmiedeke, \'{A}lvaro S\'{a}nchez-Monge, and Alexander Zernickel for intensive testing and helpful suggestions. We acknowledge funding from BMBF/Verbundforschung through the Projects ALMA-ARC 05A11PK3 and  05A14PK1.\\
    \end{acknowledgements}


    \newpage
    \clearpage
    \onecolumn
    \appendix
    \section{Examples}

    \subsection{ListDatabase function}\label{app:listdatabase}

This function reads in entries from the table \texttt{transitions}, see Sect.~\ref{sec:db}, located in the SQLite3 database file and prints out the contents (name of molecule, quantum number for lower and upper state, transition frequency, uncertainty of transition frequency, Einstein A coefficient, and energy of lower state) to the screen or file (defined by the input parameter \texttt{OutputDevice}). The user can limit the output by defining a minimum and maximum for the frequency (or for the lower energy) for the transitions or by selecting one or more molecule(s) by using the input parameter \texttt{SelectMolecule}.\\

\noindent Input within CASA:

    \begin{verbatim}
FreqMin = 20000.0
FreqMax = 990000.0
ElowMin = 100.0
ElowMax = 2000.0
SelectMolecule = ["CO;v=0;"]
OutputDevice = ""
Contents = ListDatabase()
    \end{verbatim}

\noindent Screen output (shortened):

    \begin{verbatim}
Name:   upper state:  lower state:  Freq. [MHz]:  Error Freq. [MHz]:  Einstein A[s-1]:  E_low [K]:
CO;v=0;   X;v=0;J=7;    X;v=0;J=6;  806651.80600          5.0000e-03         3.422e-05     115.354
CO;v=0;   X;v=0;J=8;    X;v=0;J=7;  921799.70000          5.0000e-03         5.134e-05     153.798
    \end{verbatim}

    \subsection{myXCLASS function}\label{app:myxclass}

The function calculates a spectrum based on a user defined molfit file described in Sect.~\ref{myxclass:molfit}. The function returns the modeled spectrum as well as the intensities and optical depths for each component.

In the example described below, we use the molfit file \texttt{CH3OH\_\_pure.molfit} located in \texttt{"demo/myXCLASS/"} to create a spectrum between 580102.0 and 580546.5~MHz with a step size of 0.5~MHz. Additionally, we use global definitions for $T_{\rm bg}$, $T_{\rm slope}$, $N_H$, $\kappa_{\nu_{\rm ref}}$, and $\beta$ (indicated by \texttt{t\_back\_flag = T} and \texttt{nH\_flag = T}). Furthermore, we use the iso ratio file \texttt{iso\_names.txt} and set the diameter of the telescope to \texttt{3.5}~m. In addition, we set the rest frequency to \texttt{0.0}~MHz with a reference velocity of 0~km s$^{-1}$. The output parameter \texttt{modeldata} contains the modeled spectrum described by a python array with two columns, where the first column indicates the frequencies (in MHz) and the second column the corresponding intensities (in K). For a rest frequency unequal zero, the myXCLASS function adds a further column to parameter \texttt{modeldata} describing the velocities (in km s$^{-1}$).

The output parameter \texttt{log} contains some informations about the calculation and parameter \texttt{trans} describes the transition frequencies of all molecules used in the molfit file within the given frequency range. The intensities and optical depths for each component and molecule is stored in parameter \texttt{IntOptical}. Finally, parameter \texttt{jobDir} indicates the path and name of the corresponding job directory containing all files produced during the calculation process. We note that the user is free to define other names for the output parameters.\\

\noindent Input within CASA:

    \begin{verbatim}
FreqMin = 580102.0
FreqMax = 580546.5
FreqStep = 5.0000000000E-01
TelescopeSize = 3.5
Inter_Flag = F
t_back_flag = T
tBack = 1.1
tslope = 0.0000000000E+00
nH_flag = T
N_H = 3.0000000000E+24
beta_dust = 2.0
kappa_1300 = 0.02
MolfitsFileName = "demo/myXCLASS/CH3OH__pure.molfit"
iso_flag = T
IsoTableFileName = "demo/myXCLASS/iso_names.txt"
RestFreq = 0.0
vLSR = 0.0
modeldata, log, TransEnergies, IntOptical, jobDir = myXCLASS()
    \end{verbatim}

    \subsection{MAGIX function}\label{app:magix}

The function starts the \texttt{MAGIX} program. The user has to define the paths and names of the registration, instance, experimental, and algorithm xml file, respectively. The parameter \texttt{MAGIXOption} indicates if informations are printed to screen (\texttt{None}) or not (\texttt{quiet}). The results, that is, the optimized instance file(s) etc., are stored in a job directory described by parameter \texttt{JobDir}, mentioned above.\\

\noindent Input within CASA:

    \begin{verbatim}
MAGIXExpXML = "demo/MAGIX/TwoOscillators_RefFit_R.xml"
MAGIXInstanceXML = "demo/MAGIX/parameters.xml"
MAGIXFitXML = "demo/MAGIX/Levenberg-Marquardt_Parameters.xml"
MAGIXRegXML = "Fit-Functions/Drude-Lorentz_conv/xml/"
MAGIXRegXML += "Conventional_Drude-Lorentz.xml"
MAGIXOption = ""
JobDir = MAGIX()
    \end{verbatim}

    \subsection{myXCLASSFit function}\label{app:myxclassfit}

This function provides a simplified interface for \texttt{MAGIX} using the myXCLASS program. The function returns the contents of the molfit file containing the parameters for the best fit and the corresponding model function values for each data point.

Here we fit observational data stored in file \texttt{band1b.dat} using the molfit file \texttt{CH3OH\_\_old.molfit}, see below, without isotopologues (indicated by \texttt{iso\_flag = F}) and the Levenberg-Marquardt algorithm with maximum ten iterations.\\

\noindent Input within CASA:

    \begin{verbatim}
NumberIteration = 10
MolfitsFileName = "demo/myXCLASSFit/CH3OH__old.molfit"
experimentalData = "demo/myXCLASSFit/band1b.dat"
TelescopeSize = 3.5
Inter_Flag = F
t_back_flag = T
tBack = 1.1
tslope = 0.0000000000E+00
nH_flag = T
N_H = 3.0000000000E+24
beta_dust = 2.0
kappa_1300 = 0.02
iso_flag = F
IsoTableFileName = ""
RestFreq = 0.0
vLSR = 0.0
newmolfit, modeldata, JobDir = myXCLASSFit()
    \end{verbatim}

\noindent Extended molfit file \texttt{CH3OH\_\_old.molfit} used in the example described above:

    \begin{small}
    \begin{verbatim}
CH3OH;v=0;   1
% limit:  s_size:   limit:    T_rot:   limit:     N_tot:   limit:  V_width:   limit:  V_off:   CFFlag:
   10.00    1.153    50.00  115.1219    50.00  1.648e+17    10.00     7.463     0.00  -7.900   c
    \end{verbatim}
    \end{small}

    \subsection{myXCLASSMapFit function}\label{app:myxclassmapfit}

The \texttt{myXCLASSMapFit} function uses the \texttt{myXCLASS} program in conjunction with \texttt{MAGIX} to fit complete data cubes instead of single spectra. In the example described below, we fit a part of the FITS cube \texttt{Orion.methanol.cbc.contsub.image.fits} described by the region file \texttt{region\_\_box\_\_ds9.reg}. (In order to fit the whole data cube, the parameter \texttt{regionFileName} has to be cleared, that is, \texttt{regionFileName = ""}.) Here, we set the threshold to two, that is, we fit all pixel which have a maximum intensity of 0.001~K (or higher). Additionally, we use the molfit file \texttt{CH3OH.molfit} together with the iso ratio file \texttt{iso\_names.txt} and set the diameter of the telescope to \texttt{3.5}~m.  Furthermore, we use the Levenberg-Marquardt algorithm with maximum 20 iterations for each pixel fit. In order to use another algorithm (or algorithm combination) the parameter \texttt{AlgorithmXMLFile} has to describe the path and name of a MAGIX algorithm XML file.
The \texttt{myXCLASSMapFit} function returns the path and the name of the job directory described by parameter \texttt{JobDir} containing subdirectories for each pixel fit and the parameter FITS images.

\newpage

\noindent Input within CASA:

    \begin{verbatim}
NumberIteration = 15
AlgorithmXMLFile = "demo/myXCLASSMapFit/algorithm-settings.xml"
MolfitsFileName = "demo/myXCLASSMapFit/CH3OH__new.molfit"
experimentalData = "demo/myXCLASSMapFit/Orion.methanol.cbc.contsub.image.fits"
regionFileName = "demo/myXCLASSMapFit/ds9_phys.reg"

FreqMin = 0.0
FreqMax = 0.0
TelescopeSize = 3.5
Inter_Flag = F
t_back_flag = T
tBack = 9.5000000000E-01
tslope = 0.0000000000E+00
nH_flag = T
N_H = 3.0000000000E+24
beta_dust = 1.4
kappa_1300 = 0.02
iso_flag = F
IsoTableFileName = ""
RestFreq = 0.0
vLSR = 0.0
Threshold = 0.001
UsePreviousResults = T
clusterdef = "demo/myXCLASSMapFit/clusterdef.txt"
JobDir = myXCLASSMapFit()
    \end{verbatim}

    \subsection{myXCLASSMapRedoFit function}\label{app:myxclassmapredofit}

This function offers the possibility to improve the result of a previous performed \texttt{myXCLASSMapFit} run using another molfit file and/or other algorithms.

In the example given below, the \texttt{myXCLASSMapRedoFit} function will re-fit the pixels \texttt{(83.201,-15.1345)} and \texttt{(83.111,-15.221)} of the previous \texttt{myXCLASSMapFit} run with job number \texttt{1234} using the molfit file \texttt{"CH3OH.molfit"}.\\

\noindent Input within CASA:

    \begin{verbatim}
JobNumber = 1234
PixelList = [[83.201, -15.1345], [83.111, -15.221]]
NumberIteration = 10
AlgorithmXMLFile = ""
MolfitsFileName = "demo/myXCLASSMapFit/CH3OH.molfit"
experimentalData = ""
Threshold = 0.0
FreqMin = 0.0
FreqMax = 0.0
t_back_flag = T
tBack = 9.5000000000E-01
tslope = 0.0000000000E+00
nH_flag = T
N_H = 3.0000000000E+24
beta_dust = 2.0
kappa_1300 = 0.02
iso_flag = T
IsoTableFileName = "demo/myXCLASSMapFit/iso_names.txt"
RestFreq = 0.0
vLSR = 0.0
myXCLASSMapRedoFit()
    \end{verbatim}
\newpage

    \section{Derivations}


    \subsection{Detection equation}\label{deriv:DetectionEquation}

In absence of scattering, the radiative transfer equation
\begin{equation}
    \frac{\mathrm{d} I_{\nu}}{\mathrm{d} s} = -\kappa_{\nu}(s) I_{\nu} + \epsilon_{\nu}(s)
\end{equation}
describes the propagation of radiation which passes through a medium. During the propagation photons are absorbed and emitted indicated by the emission and absorption coefficients $\epsilon_\nu$ and $\kappa_{\nu}$, respectively.\\

\noindent The optical depth $\tau_{\nu}(s)$ which measures the distance in units of the mean free path, is given by
\begin{equation}\label{DE:tau}
    \tau_{\nu}(s) = \int_{s'=0}^{s'=s} \kappa_{\nu} \, d s'.
\end{equation}
By using the source function $S_{\nu} = \frac{\epsilon_{\nu}}{\kappa_{\nu}}$, the radiative transfer equation can be expressed as
\begin{equation}\label{DE:dI}
    \frac{\mathrm{d} I_{\nu}}{\mathrm{d} \tau} = I_{\nu}(\tau) + S_{\nu}(\tau).
\end{equation}
Within the local thermodynamic equilibrium (LTE) the source function is described by Planck's law for black body radiation $B_{\nu}(T)$.\\

\noindent Integrating Eq.~\eqref{DE:dI} over $\tau$ leads to
\begin{equation}
    I_{\nu}(s) = I_{\nu}(0) \, e^{-\tau_{\nu}} + \int_{\tau'=0}^{\tau'=\tau} S_{\nu}(\tau'_{\nu}) e^{\tau'_{\nu} - \tau_{\nu}} d \tau'_{\nu}.
\end{equation}
Assuming a constant source function, that is, constant emission and absorption coefficients through the medium, the transfer equation can be written as
\begin{equation}\label{DE:final}
    I_{\nu}(s) = I_{\nu}(0) \, e^{-\tau_{\nu}} + S_{\nu} \left(1 - e^{-\tau_{\nu}} \right).
\end{equation}
Additionally, we have to take into account, that the different components may not cover the whole beam, that is, that the background behind a certain component might contribute as well. So, we have to extend Eq.~\eqref{DE:final} by introducing the beam filling factor $\eta$, Eq.~\eqref{myXCLASS:BeamFillingFactor}, which describes the fraction of the beam covert by a component
\begin{equation}\label{DE:finalExt}
    I_{\nu}(s) = \eta \, \left[I_{\nu}(0) \, e^{-\tau_{\nu}} + S_{\nu} \left(1 - e^{-\tau_{\nu}} \right)\right] + \left(1 - \eta\right) \, I_{\nu}(0).
\end{equation}
Here, the term $\eta \, I_{\nu}(0) \, e^{-\tau_{\nu}}$ indicates the attenuated radiation from the background $I_{\nu}(0)$. The second term $\eta \, S_{\nu} \left(1 - e^{-\tau_{\nu}} \right)$ describes the self attenuated radiation emitted by a certain component. Finally, the last term $\left(1 - \eta\right) \, I_{\nu}(0)$ represents the contribution of the background which is not covert by a component.\\

In real observations, we do not measure absolute intensities but only differences of intensities, that is, we have to subtract the OFF-position for single dish observations from Eq.~\eqref{DE:finalExt} as well, where we have an intensity caused by the cosmic background $J_\mathrm{CMB}$. So, we achieve
\begin{align}\label{DE:FinalEq}
    I_{\nu}(s) &= \eta \, \left[I_{\nu}(0) \, e^{-\tau_{\nu}} + S_{\nu} \left(1 - e^{-\tau_{\nu}} \right)\right] + \left(1 - \eta\right) \, I_{\nu}(0) - J_\mathrm{CMB}\nonumber\\
               &= \eta \, \left[\left(S_{\nu} - I_{\nu}(0)\right)\left(1 - e^{-\tau_{\nu}} \right)\right] + I_{\nu}(0) - J_\mathrm{CMB}.
\end{align}
%


    \subsection{Beam filling factor}\label{deriv:BeamFillingFactor}

    \begin{figure}[t]
       \centering
       \includegraphics[width=0.69\textwidth]{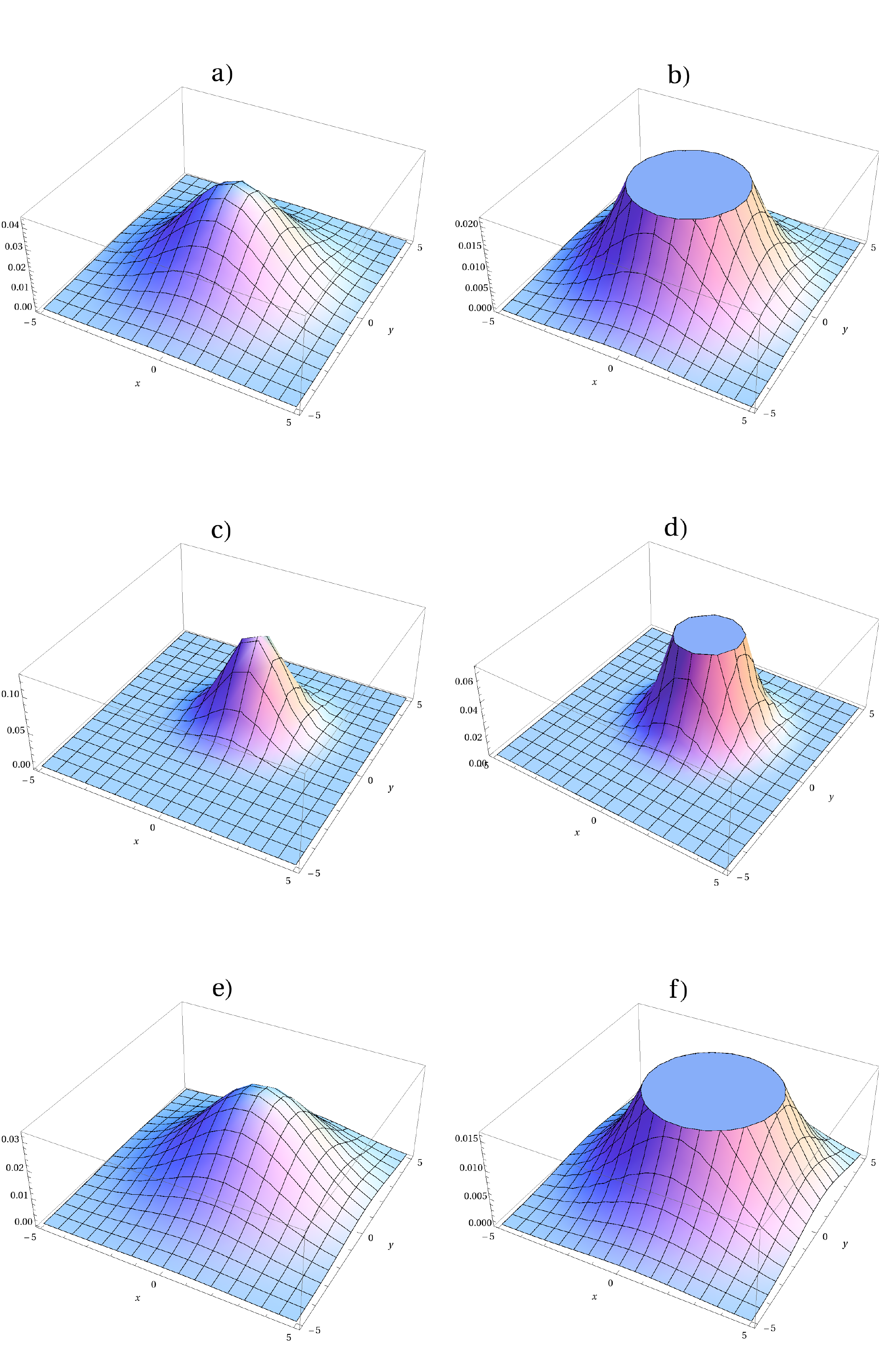}\\
       \caption{a)~A single two-dimensional Gaussian function, Eq.~\eqref{bFf:General2DGauss}, with $\sigma_x = \sigma_y = 1.9$ and $\mu_x = \mu_y = 0$ representing a Gaussian beam of a telescope. b)~Cut through the Gaussian function described in a) at half height. c)~A single two-dimensional Gaussian function, with $\sigma_x = \sigma_y = 1.1$ and $\mu_x = 0.5$ and $\mu_y = 0.9$ representing a Gaussian beam of a point source. d)~Cut through the Gaussian function described in c) at half height. e)~Convolved Gaussian of telescope and point source as given by Eq.~\eqref{bFf:GaussConvSimp}. f)~Cut through the convolved Gaussian function described in e) at half height.}
       \label{fig:bFf}
    \end{figure}

The derivation of the beam filling factor \eqref{myXCLASS:BeamFillingFactor} starts with the normalized two-dimensional Gaussian function
\begin{equation}\label{bFf:General2DGauss}
    g(x, y, \sigma_x, \sigma_y, \mu_x, \mu_y) = \frac{1}{\sqrt{2 \, \pi \, \left(\sigma_x^2+\sigma_y^2\right)}} \, e^{-\left(\frac{\left(x-\mu_x\right)^2}{2 \, \sigma_x^2}+\frac{\left(y - \mu_y\right)^2}{2 \, \sigma_y^2}\right)},
\end{equation}
where $\sigma_x^2$ and $\sigma_y^2$ describe the variances and $\mu_x$ and $\mu_y$ the center along the $x$ and $y$ axis, respectively.\\

Observing a Gaussian shaped extended source with a telescope is described by a convolution of two two-dimensional Gaussian functions:
\begin{align}\label{bFf:GaussConvGeneral}
    \left(g_1 * g_2 \right) &= \int_{-\infty}^\infty \, \int_{-\infty}^\infty \, g_1(x - u, y - v, \sigma_{x,1}, \sigma_{y,1}, \mu_{x, 1}, \mu_{y, 1}) \cdot g_2(u, v, \sigma_{x,2}, \sigma_{y,2}, \mu_{x,2}, \mu_{y, 2}) \, du \, dv \nonumber\\
                            &= \frac{1}{2 \pi  \sqrt{\left(\sigma_{x,1}^2 + \sigma_{x,2}^2\right) \left(\sigma_{y,1}^2 + \sigma_{y,2}^2\right)}} \, e^{-\left(\frac{\left(\mu_{x,1} + \mu_{x,2} - x\right)^2}{2 \, \left(\sigma_{x,1}^2 + \sigma_{x,2}^2\right)} + \frac{\left(\mu_{y,1} + \mu_{y,2} - y\right)^2}{2 \, \left(\sigma_{y,1}^2 + \sigma_{ y,2}^2\right)}\right)}.
\end{align}
Assuming that $g_1$ describes the telescope with $\mu_{x,1} = \mu_{y, 1} \equiv 0$ and that telescope and extended source are described by non-elliptical Gaussians, that is, $\sigma_{x,1} = \sigma_{y,1} \equiv \sigma_1$ and $\sigma_{x,2} = \sigma_{y,2} \equiv \sigma_2$, Eq.~\eqref{bFf:GaussConvGeneral} can be simplified to\footnote{Here ``*'' indicates the convolution of two functions $g_1$ and $g_2$.}
\begin{equation}\label{bFf:GaussConvSimp}
    \left(g_1 * g_2 \right) = \frac{1}{2 \pi \left(\sigma_1^2 + \sigma_2^2\right)} \, e^{-\frac{\left(x - \mu_{x,2}\right)^2 + \left(y - \mu_{y,2}\right)^2}{2 \left(\sigma_1^2  + \sigma_2^2\right)}}.
\end{equation}
The FWHM of the resulting Gaussian is given by
\begin{equation}\label{bFf:GaussConvSimpFWHM}
    {\rm FWHM} = 2 \sqrt{2 \, \log 2} \sqrt{\sigma_1^2 + \sigma_2^2}
\end{equation}
which describe an area of
\begin{align}\label{bFf:EllipseArea}
    A_{\rm conv}^{\rm FWHM} &= \pi \cdot 2 \, \log 2 \cdot \left(\sigma_1^2 + \sigma_2^2\right) \nonumber\\
                 &= \frac{\pi}{4} \cdot \left(\theta_1^2 + \theta_2^2\right).
\end{align}
In the last line we used the relation between the variances $\sigma_{1,2}$ and the user defined FWHM of telescope ($\theta_1 \equiv \theta_t$) and source size ($\theta_2 \equiv \theta^{m,c}$) which is given by
\begin{equation}\label{bFf:FWHMvariance}
    \theta_i = 2 \, \sqrt{2 \, \log 2} \cdot \sigma_i.
\end{equation}
The beam filling factor Eq.~\eqref{myXCLASS:BeamFillingFactor} is defined as ratios of areas
\begin{equation}\label{bFf:etaGeneral}
    \eta_{g_1,g_2} = \frac{A_{\rm source}^{\rm FWHM}}{A_{\rm conv}^{\rm FWHM}} = \frac{\frac{\pi \, \theta_2^2}{4}}{\frac{\pi}{4} \cdot \left(\theta_1^2 + \theta_2^2\right)} = \frac{\theta_2^2}{\left(\theta_1^2 + \theta_2^2\right)},
\end{equation}
which is completely independent of the position $\mu_x$ and $\mu_y$ of the extended source within the telescope beam.\\

If more extended sources are observed with the telescope, we have to convolve the already convolved Gaussian Eq.~\eqref{bFf:GaussConvSimp} with a further two-dimensional Gaussian function with variance $\sigma_{x,3} = \sigma_{y,3} \equiv \sigma_3$ and center $\mu_{x,3}$ and $\mu_{y,3}$. We get
\begin{align}
  \left(\left(g_1 * g_2 \right) * g_3 \right) = \left(g_1 * g_2 * g_3 \right) = \frac{1}{2 \pi \left(\sigma_1^2 + \sigma_2^2 + \sigma_3^2\right)} \, e^{\left(-\frac{\left(\mu_{x,2} + \mu_{x,3} - x\right)^2 + \left(\mu_{y,2} + \mu_{y,3} - y\right)^2}{2 \, \left(\sigma_1^2 + \sigma_2^2 + \sigma_3^2\right)}\right)}.
\end{align}
The FWHM of the resulting Gaussian is given by
\begin{align}\label{bFf:NextGaussConvSimpFWHM}
    {\rm FWHM} = 2 \sqrt{2 \, \log 2} \sqrt{\sigma_1^2 + \sigma_2^2 + \sigma_3^2}
\end{align}
which describe an area of
\begin{align}\label{bFf:NextEllipseArea}
    A_{\rm conv}^{\rm next} &= \pi \cdot 2 \, \log 2 \cdot \left(\sigma_1^2 + \sigma_2^2 + \sigma_3^2\right) \nonumber\\
                 &= \frac{\pi}{4} \cdot \left(\theta_1^2 + \theta_2^2 + \theta_3^2\right).
\end{align}
In the last line we used again the relation between the variances and FWHM, Eq.~\eqref{bFf:FWHMvariance}. Now, we can again define a beam filling factor similar to Eq.~\eqref{myXCLASS:BeamFillingFactor} and get
\begin{equation}\label{bFf:NextEtaGeneral}
    \eta_{g_1,g_2,g_3} = \frac{\frac{\pi \, \theta_2^2}{4}+\frac{\pi \, \theta_3^2}{4}}{\frac{\pi}{4} \cdot \left(\theta_1^2 + \theta_2^2 + \theta_3^2\right)} = \frac{\theta_2^2+\theta_3^2}{\left(\theta_1^2 + \theta_2^2 + \theta_3^2\right)}.
\end{equation}
When we observe more than three extended sources with a telescope we have to iteratively convolve the resulting Gaussian with a further two-dimensional Gaussian function and we can generalize Eq.~\eqref{bFf:NextEtaGeneral} to
\begin{equation}\label{bFf:NextNextEtaGeneral}
    \eta_{\rm general} = \frac{\sum_{i > 1} \theta_i^2}{\left(\sum_i \theta_i^2\right)} = \frac{\sum_{i > 1} \theta_i^2}{\theta_t^2 + \sum_{i > 1} \theta_i^2},
\end{equation}
where $\theta_1 \equiv \theta_t$ describes the FWHM of the Gaussian shaped beam of the telescope, defined by the diffraction limit, Eq.~\eqref{myXCLASS:DiffractionLimit}.


    \subsection{Optical depth}\label{deriv:OpticalDepth}

    \begin{figure}[t]
       \centering
       \includegraphics[width=0.3\textwidth]{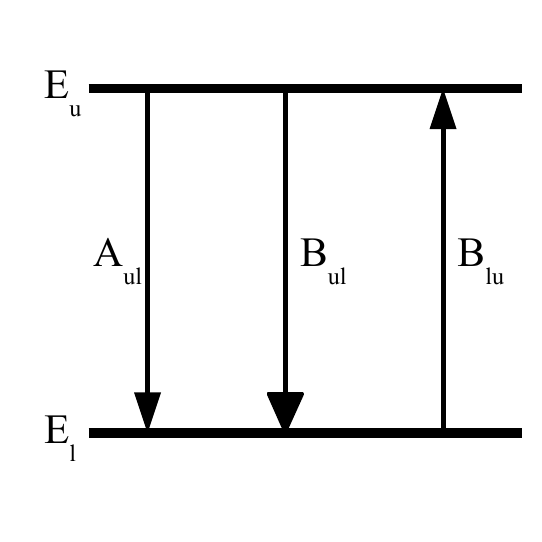}\\
       \caption{Transitions between the lower $l$ and the upper $u$ level with the corresponding Einstein coefficients.}
       \label{fig:od}
    \end{figure}

In order to derive Eq.~\eqref{myXCLASS:tau} we consider a system which involves radiative transitions between a lower $l$ and an upper $u$ level only. As shown in Fig.~\ref{fig:od}, the lower level has an energy $E_l$ and the upper level an energy $E_u > E_l$. With
\begin{equation}\label{od:nu}
    h \, \nu_{u,l} = E_u - E_l
\end{equation}
describing the energy difference between these two levels we can express the emissivity due to spontaneous radiative decay as
\begin{equation}\label{od:epsilon}
    \epsilon_{l,u,\nu} = \frac{h \, \nu}{4 \, \pi} \, n_u \, A_{u,l} \, \phi_{l,u} (\nu),
\end{equation}
where $A_{u,l}$ describes the Einstein A-coefficient, or radiative decay rate for the transition from the lower~$l$ to the upper~$u$ level. The expression $A_{u,l}^{-1}$ gives the averaged time, that a quantum mechanical system can stay in level~$u$ before radiatively decaying to level~$l$, where we assume no collisional (de-)excitation. The expression $\phi_{l,u} (\nu)$ describes the line profile of the transition of photons of frequency $\nu$ and is normalized to one, that is, $\int_0^\infty \phi(\nu) \, d \nu = 1$.\\

Similar to Eq.~\eqref{od:epsilon} we can write the extinction coefficient, which describes the radiative excitation from the lower to the upper level
\begin{equation}\label{od:kappaExt}
    \kappa_{l,u,\nu}^{\rm ext} = \frac{h \, \nu}{4 \, \pi} \, n_l \, B_{l,u} \, \phi_{l,u} (\nu),
\end{equation}
where $B_{l,u}$ describes the Einstein B-coefficient for extinction.\\

In addition to spontaneous emission and extinction we have to take the stimulated emission into account, which can be described by adding a negative opacity contribution to Eq.~\eqref{od:kappaExt}:
\begin{equation}\label{od:kappa}
    \kappa_{l,u,\nu} = \frac{h \, \nu}{4 \, \pi} \, \left(n_l \, B_{l,u} - n_u \, B_{u,l}\right) \, \phi_{l,u} (\nu),
\end{equation}
where $B_{u,l}$ represents the Einstein B-coefficient for stimulated emission. So, for $n_l B_{l,u} < n_u \, B_{u,l}$ we get laser (maser) emission.\\

The different Einstein coefficients are related to each other by the Einstein relations:
\begin{align}\label{od:EinsteinRelations}
    A_{u,l} &= \frac{2 \, h \, \nu^3}{c_{\rm light}^2} \, B_{u,l},\nonumber\\
    g_l \, B_{l,u} &= g_u \, B_{u,l}.
\end{align}
Following Eq.~\eqref{DE:tau}, the differential optical depth $\tau_{\nu}$ is defined as
\begin{align}\label{od:tauFirstPart}
    d \tau_{\nu} = \kappa_{\nu} \, ds &= \left(\frac{h \, \nu}{4 \, \pi} \, \left(n_l \, B_{l,u} - n_u \, B_{u,l}\right) \, \phi_{l,u} (\nu) \right) \, ds \nonumber\\
    &= \left(\frac{c_{\rm light}^2}{8 \, \pi \, \nu^2} \, A_{u,l} \, \left(n_l \, \frac{g_u}{g_l} - n_u \right) \, \phi_{l,u} (\nu) \right) \, ds,
\end{align}
where we used the Einstein relations Eqn.~\eqref{od:EinsteinRelations} in the last line. By assuming LTE conditions and therefore Boltzmann population distribution
\begin{equation}
    \frac{n_u}{n_l} = \frac{g_u}{g_l} \, e^{(-E_u - E_l)/k_B T_{\rm ex}} = \frac{g_u}{g_l} \, e^{-h \, \nu_{u,l}/k_B T_{\rm ex}}
\end{equation}
we can rewrite Eq.~\eqref{od:tauFirstPart} by using Eq.~\eqref{od:nu}
\begin{equation}\label{od:tau2ndPart}
    d \tau_{\nu} = \left(\frac{c_{\rm light}^2}{8 \, \pi \, \nu^2} \, A_{u,l} \, n_u \, \left(e^{h \, \nu_{u,l} / k_B \, T_{\rm ex}} - 1 \right) \, \phi_{l,u} (\nu) \right) \, ds.
\end{equation}
Finally, we have to integrate along the line of sight and obtain the optical depth $\tau_{\nu}$
\begin{equation}\label{od:tau3rdPart}
    \tau_{\nu} = \frac{c_{\rm light}^2}{8 \, \pi \, \nu^2} \, A_{u,l} \, N_u \, \left(e^{h \, \nu_{u,l} / k_B \, T} - 1 \right) \, \phi_{l,u} (\nu),
\end{equation}
where $N_u = \int n_u \, ds$ describes the column density of a certain molecule in the upper state.\\

In order to express Eq.~\eqref{od:tau3rdPart} in terms of the total column density $N_{\rm tot} = \sum_{i=0}^\infty n_i$ we start again with the Boltzmann population distribution
\begin{equation}\label{od:NtotGeneral}
    N_{\rm tot} = \sum_{i=0}^\infty n_i = n_j \, \sum_{i=0}^\infty \frac{n_i}{n_j} = n_j \, \sum_{i=0}^\infty \frac{g_i}{g_j} \, e^{(-E_i+E_j)/k_B T_{\rm ex}} = \frac{n_j}{g_j} \, e^{E_j/k_B T_{\rm ex}} \cdot Q \left(T_{\rm ex} \right),
\end{equation}
where we used the partition function $Q(T_{\rm ex})$, which is defined as sum over all states
\begin{equation}\label{od:PartFunc}
    Q(T_{\rm ex}) = \sum_{i=0}^\infty g_i \, e^{-E_i/k_B \, T_{\rm ex}}.
\end{equation}
For our purpose we rewrite Eq.~\eqref{od:NtotGeneral} in terms of $N_u$ and $N_{\rm tot}$:
\begin{equation}\label{od:NtotUL}
    N_{\rm tot} = \frac{N_u}{g_u} \, e^{E_u/k_B T_{\rm ex}} \cdot Q \left(T_{\rm ex} \right).
\end{equation}
Inserting this expression into Eq.~\eqref{od:tau3rdPart} gives
\begin{align}\label{od:tauFinal}
    \tau_{\nu} &= \frac{c_{\rm light}^2}{8 \, \pi \, \nu^2} \, A_{u,l} \, \left[\frac{N_{\rm tot} \, g_u}{Q \left(T_{\rm ex} \right)} \, e^{-E_u/k_B T_{\rm ex}} \right] \, \left(e^{h \, \nu_{u,l} / k_B \, T} - 1 \right) \, \phi_{l,u} (\nu) \nonumber\\
    &= \frac{c_{\rm light}^2}{8 \, \pi \, \nu^2} \, A_{u,l} \, N_{\rm tot} \, \frac{g_u \, e^{-E_l/k_B T_{\rm ex}}}{Q \left(T_{\rm ex} \right)} \, \left(1 - e^{-h \, \nu_{u,l} / k_B \, T} \right) \, \phi_{l,u} (\nu),
\end{align}
where we used Eq.~\eqref{od:nu} to achieve Eq.~\eqref{myXCLASS:tau} for a single line of a certain component and molecule.

\end{document}